\documentclass[12pt, a4paper]{article}

\usepackage{amsmath,amssymb}
\usepackage{tabularx}

\usepackage{changepage}

\usepackage{textcomp,marvosym}

\usepackage{cite}

\usepackage{nameref,hyperref}

\usepackage[nopatch=eqnum]{microtype}
\DisableLigatures[f]{encoding = *, family = * }

\usepackage[table]{xcolor}

\usepackage{array}

\usepackage{graphicx}
\usepackage[aboveskip=1pt,labelfont=bf,labelsep=period,justification=raggedright,singlelinecheck=off]{caption}

\usepackage{vmargin}
\setmargins{2.0cm}{2.0cm}{17.0 cm}{24cm}{0.5cm}{0cm}{1cm}{1cm}

\begin{document}
\vspace*{0.2in}

\begin{flushleft}
{\Large
\textbf\newline{Dynamics of temporal influence in polarised networks} 
}
\newline
\\
Caroline B. Pena\textsuperscript{1*},
David J.P. O'Sullivan\textsuperscript{1},
Pádraig MacCarron\textsuperscript{1},
Akrati Saxena\textsuperscript{2}
\\
\bigskip
\textbf{1} Mathematics Applications Consortium for Science and Industry (MACSI), Department of Mathematics \& Statistics, University of Limerick, Limerick V94T9PX, Ireland
\\
\textbf{2} Leiden Institute of Advanced Computer Science (LIACS), Leiden University, 2300 RA Leiden, The Netherlands
\\
\bigskip

* caroline.pena@ul.ie

\end{flushleft}
\section*{Abstract}
In social networks, it is often of interest to identify the most influential users who can successfully spread information to others. This is particularly important for marketing (e.g., targeting influencers for a marketing campaign) and to understand the dynamics of information diffusion (e.g., who is the most central user in the spreading of a certain type of information). However, different opinions often split the audience and make the network polarised. In polarised networks, information becomes soiled within communities in the network, and the most influential user within a network might not be the most influential across all communities.
Additionally, influential users and their influence may change over time as users may change their opinion or choose to decrease or halt their engagement on the subject. In this work, we aim to study the temporal dynamics of users' influence in a polarised social network. We compare the stability of influence ranking using temporal centrality measures, while extending them to account for community structure across a number of network evolution behaviours. We show that we can successfully aggregate nodes into influence bands, and how to aggregate centrality scores to analyse the influence of communities over time. A modified version of the temporal independent cascade model and the temporal degree centrality perform the best in this setting, as they are able to reliably isolate nodes into their bands.

\section{Introduction}
Information spread plays an important role in shaping people's opinions and behaviour in social networks~\cite{ausat2023role,swastiningsih2024role}. Currently, information spread is faster and easier than in the past, with the use of online social media where sharing information with connections is just one click away. Online social platforms, such as Facebook, Instagram, TikTok, and Twitter (currently known as X), serve as a venue for the information spread among their users, where users both create and share content with each other~\cite{amedie_impact_2015, allcott_social_2017}. Understanding how information spreads on social networks is of paramount importance for society~\cite{amedie_impact_2015, siddiqui2016social}, having applications in public health~\cite{clark_role_2022, johnson_online_2020, beguerisse-diaz_who_2017}, politics~\cite{amedie_impact_2015, persily_social_2020}, and business~\cite{aral_distinguishing_2009}. Information spread in social networks is commonly influenced by homophily, i.e., people's tendency to associate preferentially with other people who are similar to themselves in some way~\cite{newman2003structure, rizi2024homophily}. Users of social media platforms have a tendency to group with others that share similar opinions and interests, and tend to share information from those who are similar~\cite{pena2025finding, o2017integrating, kearney2019analyzing, chen2021polarization}. The division of society into groups that believe in different, often opposing, ideas is commonly referred to as polarisation~\cite{smith2024polarization}, which is also observed in social media discussions, especially where the topic is controversial. 

Online social platforms, such as X (formerly known as Twitter), Facebook, and others, further amplify polarisation among users by using a self-reinforcing system where users are more likely to see the posts from others they share opinions with~\cite{smith2024polarization, flamino2023political, darwish2019quantifying, chen2021polarization}. O'Sullivan et al.~\cite{o2017integrating} and Pena et al.~\cite{pena2025finding} explored the polarisation structure on conversation networks on Twitter about two recent referendums in Ireland: (i) the same-sex marriage referendum of 2015, and (ii) the abortion referendum of 2018. Both studies showed that users involved in the online conversation around these referendums presented a strong homophilic behaviour, leading to the observed polarisation. Kearney~\cite{kearney2019analyzing} studied network polarization on Twitter during the 2016 general election in the USA, and also observed that partisan users form highly polarised networks, while moderates and less engaged users largely avoid political discussions. Researchers have also studied the evolution of polarisation and its impact on opinion formation. De Arruda et al.~\cite{de2022modelling} modelled opinion dynamics in online social networks, showing that friendship rewiring and network algorithms influence polarisation and echo chamber formation, and the temporal dynamics can lead to scenarios ranging from consensus to extreme polarisation. Soares et al.~\cite{soares2018influencers} analysed influencers' roles in political conversations on Twitter during the impeachment process of the ex-president of Brazil, Dilma Rousseff. The authors observed that the network is highly modularised and contains three types of influencers shaping influence and polarisation --- opinion leaders, informational influencers, and activists. Loy et al.~\cite{loy2022opinion} proposed a Boltzmann-type kinetic model for opinion formation in social networks, considering connectivity-based opinion influence. There are several other works~\cite{baumann2020modeling, cinelli2021echo, jiang2021social}, which have studied the evolution of echo-chambers in polarised networks and observed that information tends to flow within its own group.

In such highly opinionated environments, community structure serves as good indicators of polarised groups~\cite{pena2025finding, o2017integrating}. Consequently, when analysing such complex and polarised social networks, it is of interest to identify the most influential nodes within each polarised community over time. These central ``players" drive information spread by convincing others to share their content or news within their connections. We can measure the influential power of a user using different centrality measures~\cite{newman2018networks}. In the literature, several centrality measures have been defined, which are used extensively to identify influential nodes who maximize the influence spread, i.e., if they start sharing the content on the network they would be expected to have a larger outreach than other nodes~\cite{saxena2020centrality, soares2018influencers, loy2022opinion}. Centrality measures have several other applications, including finding the source of rumours~\cite{shelke2019source, dang2016toward, azzimonti2023social}, identifying weak points in the network (where if nodes are removed, the structural properties of the network would deteriorate), or which nodes, if added, would improve infrastructure~\cite{pitts1965graph, 
cadini2008using, oliva2019aggregating, kaur2016analyzing, latora2005vulnerability}, and for organizational design~\cite{bizzi2024origin, beauchamp1965improved}. In this paper, we use centrality measures to identify the most influential users in polarised temporal social networks~\cite{saxena2020centrality, peng2018influence, zhang2017degree, sun2021community, landherr2010critical}. Some of the well-known centrality measures in social network analysis are~\cite{landherr2010critical}: degree centrality, closeness centrality~\cite{bavelas1950communication, koschutzki2005centrality}, betweenness centrality~\cite{freeman1977set}, eigenvector centrality~\cite{seeley1949net, bonacich1972factoring}, Katz centrality~\cite{katz1953new}, and PageRank centrality~\cite{brin1998anatomy}. There have been proposed methods to update these centrality measures in networks with communities~\cite{ghalmane2019centrality, rajeh2021characterizing} as well as extend them for temporal networks~\cite{kim2012temporal, taylor2017eigenvector, grindrod2011communicability, tsalouchidou2020temporal, tang2010analysing, alsayed2015betweenness}. However, to the best knowledge of the authors, the literature is scarce when it comes to the study of centrality on temporal networks with community structure.  

Ghalmane et al.~\cite{ghalmane2019centrality} and Rajeh et al.~\cite{rajeh2021characterizing} conducted an extensive analysis into how centrality can be calculated on networks with communities. However, they have focused on static networks with no temporal component. In real-world, diffusion mechanisms commonly unfold in a given time frame, where information takes time to spread around nodes in the network. For example, Holme~\cite{holme2016temporal} has investigated disease spreading over time on empirical datasets of human contacts; Goel et al.~\cite{goel2016structural} have analysed the virality of information in social media through a mechanistic model that infers the paths of diffusion by bringing time information into play; and Kim an Anderson~\cite{kim2012temporal} analysed the temporal dynamics of contact traces of mobile devices owned by students and staff in two universities. Therefore, it is essential to investigate how information spreads over time and identify the most influential nodes at each defined time step (e.g., hours, days, or intervals between significant events). Additionally, understanding the impact of community structures on information diffusion within these temporal networks remains a key area of interest.

In addition to the natural temporal aspect of social networks, \cite{soares2018influencers} has shown that users tend to cluster based on their level of influence within the network. Similarly, O'Brien et al.~\cite{o2021identification} ranked users in an online game and analysed the evolution of their ranks across multiple time points throughout a season. Following this approach, we categorize nodes in our networks according to their level of influence, which we refer to as \textit{influence bands}, and analyse the flow of users between these bands over time to gain insights into the temporal evolution of influence.
This method is particularly useful because (1) a user's influence naturally fluctuates over time, and while minor position shifts may be unimportant, substantial changes --- such as moving between influence bands --- can be more meaningful, and (2) simultaneously analysing both temporal and community-based influence can be complex, whereas grouping users into influence bands provides a more structured and interpretable framework for analysis.

In this work, we aim to investigate the temporal dynamics of nodes' influence in polarised online social networks by addressing the following key questions: 

\begin{enumerate}
    \item Can influencers be effectively grouped into influence bands? 
    \item Does the overall influence of a specific community within a polarised network change over time?
    \item Can we determine which polarised community the most influential nodes belong to, and how do influential nodes differ across communities?
\end{enumerate}

In the following section, we explain the methods used to compute temporal centrality measures, as well as the generative models to build synthetic networks for our analysis.

\section{Materials and methods}

In this section, we explain three different methodologies to compute temporal centralities, and our method to generate synthetic temporal networks with bands and communities. Building synthetic networks is crucial to understanding how centrality methods perform in simple and controlled temporal polarised networks. In this section we also summarize the networks studied and explain our method to aggregate nodes into influence bands.

\subsection{Temporal centrality methods}\label{sub:model_temporal}

In order to calculate temporal centrality scores, we use different techniques to represent temporal networks, which makes it more convenient to compute different centralities. Please note that these techniques are applied to the same set of networks, but are stored in different form for faster computation of centralities during the analysis.

First, we use a method proposed by Kim and Anderson~\cite{kim2012temporal}, which allows the calculation of temporal degree, closeness and betweenness centrality scores. This method involves creating a layer for each time-slice, starting at $t=0$, and every link is drawn between time-slices; refer to Fig.~\ref{Fig1}. For eigenvector-based centralities, such as eigenvector centrality, Katz centrality and PageRank, we use a second technique proposed by Taylor et al.~\cite{taylor2017eigenvector}, which creates a multilayer network where each layer contains a time slice of the temporal network, and each node is connected to itself in the subsequent and the preceding time slices (Fig.~\ref{Fig2}~(a)). We build what the authors refer to as a supra-centrality matrix, which contains the centrality values for each time-slice's centrality in block matrix form in Eq.~\ref{eq:C_matrix}.     

\subsubsection{Temporal degree and closeness centrality}\label{sub:temporal_degree}

\begin{figure}[!ht]
\centering
\includegraphics[width = 0.8\textwidth]{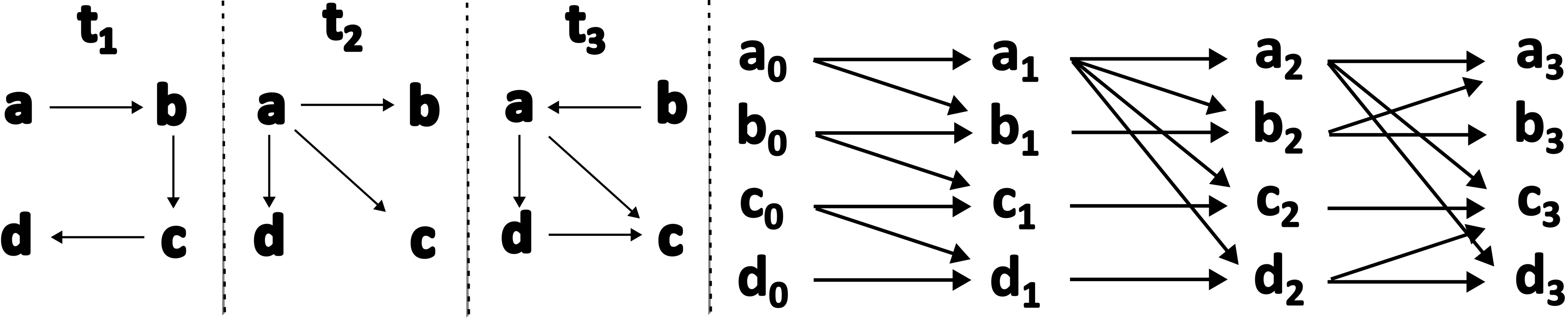}
\caption{{\bf Schematic of the network built for temporal degree and closeness centralities.} Example of a simple network analysed over three time-slices. On the left, a representation of the time-slices, and on the right, the representation of the temporal network according to the method described in~\cite{kim2012temporal}.}
\label{Fig1}
\end{figure}

Kim and Anderson~\cite{kim2012temporal} developed a method to calculate temporal degree, temporal betweenness and temporal closeness centrality scores by using a common temporal network representation (Fig.~\ref{Fig1}). The method involves creating a layer for each time step that contains a set of dummy nodes, starting at $t=0$. Each dummy node is then connected to itself in the subsequent time-slice (i.e., the dummy node $a_0$ is connected to the dummy node $a_1$, and so on), as well as to the dummy nodes they have an original connection with (e.g., if a link $a \rightarrow b$ exists in time-slice 1, the dummy connection will be written as $a_0 \rightarrow b_1$).The temporal centrality matrix looks the following.

\begin{equation}
    \mathbb{M} = 
\begin{bmatrix} 
    0 &\textbf{A}^{(1)} + \textbf{I} & 0 & 0 & 0 & ... & 0\\
    0 & 0 & \textbf{A}^{(2)} + \textbf{I} & 0 & 0 & ... & 0\\
    0 & 0 & 0 & \textbf{A}^{(3)} + \textbf{I} & 0 & ... & 0\\
    . & . & . & . & . && . \\
    . & . & . & . & . && . \\
    . & . & . & . & . && . \\
    0 & 0 & 0 & 0 & 0 & ... & \textbf{A}^{(t)} + \textbf{I}\\
    0 & 0 & 0 & 0 & 0 & ... & 0\\
\end{bmatrix},
\label{eq:M_matrix}
\end{equation}

where $\textbf{{A}}^{(t)}$ is the adjacency matrix for time step $t$ and $\textbf{I}$ add self-links between time steps. The matrix $\mathbb{M}$ is of dimension $(N \times (t+1)) \times (N \times(t+1))$, where $N$ is the number of nodes in the network and $t$ is the number of time steps.

\bigskip

For this method, a node $v$’s temporal degree is the normalised total number of inbound edges to and outbound edges from $v$ on the time interval $[i,j]$, disregarding the self-edges from $v_{t-1}$ to $v_t$ for all $t \in \{i + 1, . . . ,j\}$.

Temporal closeness requires a more complex setup. The authors define temporal closeness by considering $m$ time intervals $[t,j] : i \leq t < j$ where $m = j - i$ by varying the starting time $t$ of each time interval from $i$ to $j - 1$. The temporal shortest paths from node $u$ to node $v$ are then calculated. These are the paths from node $u_i$ to node $v_k$, which is the first node encountered along a path from $u_i$ to a node in $\{v_{i+1}, . . . ,v_j \}$. However, the temporal shortest paths from $u$ to $v$ will change as time increases. Therefore, in addition to the case with the starting time $i$, we also need to consider the temporal shortest paths from node $u$ to node $v$ on the additional $m - 1$ time intervals $[t,j] : i < t <j$ by varying $t$ from $i + 1$ to $j - 1$. A node $v \in V$ in the time interval $[i,j]$ has temporal closeness centrality calculated by

\begin{equation*}
    C_{i,j}(v) = \sum_{i \leq t < j} \sum_{u \in V \backslash v} \frac{1}{\Delta_{t,j}(v,u)},
    \label{eq:temp_closeness}
\end{equation*}
where $\Delta_{t,j}(v,u)$ is the temporal shortest path distance from $v$ to $u$ on a time interval $[t,j]$.

This way we are able to calculate temporal degree and temporal closeness for each dummy node in each time-slice. Next we explore a temporal method developed to compute eigenvector-based centrality scores.

Temporal betweenness, however, does not allow the computation of a score for each temporal dummy node due to the nature of its processing, i.e., a node $v_i$ is treated as the same as $v_k$ in another time-slice, therefore a shortest temporal path from $v$ to $u$ that looks like $v_0 \to v_1 \to u_2 \to u_3$ has no node midway to account for (no node to calculate betweenness for), as $v_0$ and $v_1$ are the same node, as well as $u_2$ and $u_3$ are the same node. For this reason, although temporal betweenness is useful for calculating the aggregate score of a node over time, it does not fit this paper's purposes.




\subsubsection{Temporal eigenvector-based centrality}\label{subsub:temporal_eigen}

\begin{figure}[!ht]
\includegraphics[width = \textwidth]{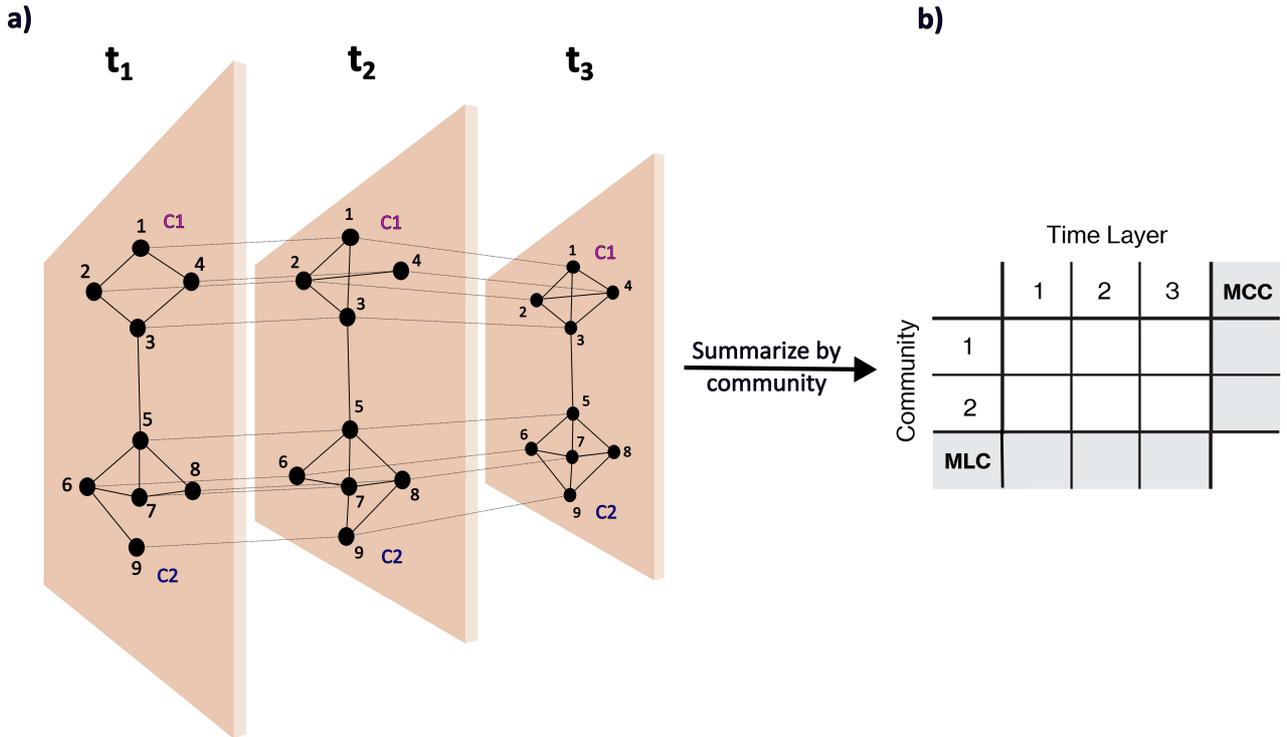}
\caption{{\bf Schematic of the multilayer network built for temporal eigenvector-based centralities for networks with communities.} Example of a network with two communities analysed over three time-slices. There is a relationship between time layers, i.e., each node is linked to itself on preceding and subsequent time layers, where the weight of each inter-layer link is $\omega = \frac{1}{\varepsilon}$. }
\label{Fig2}
\end{figure}

Taylor et al.~\cite{taylor2017eigenvector} proposed the eigenvector-based centrality for temporal networks, which extends the static eigenvector centrality by creating a multilayer network where each layer contains a time-slice of the temporal network, and each node is connected to itself in the subsequent and the preceding time-slices. Figure~\ref{Fig2}~(a) illustrates the multilayer network generated. The eigenvectors on the supra-centrality matrix are defined as:

\begin{equation}
        \mathbb{C}(\varepsilon) =
\begin{bmatrix} 
	\varepsilon \textbf{C}^{(1)} & \textbf{I} & 0 & 0 & ... & 0\\
	\textbf{I} & \varepsilon \textbf{C}^{(2)} & \textbf{I} & 0 & ... & 0\\
	0 & \textbf{I} & \varepsilon \textbf{C}^{(3)} & \textbf{I} & ... & 0\\
    . & . & . & . & & . \\
    . & . & . & . & & . \\
    . & . & . & . & & . \\
	0 & 0 & 0 & 0 & ... & \varepsilon \textbf{C}^{(t)}\\
\end{bmatrix},
\label{eq:C_matrix}
\end{equation}

where $\textbf{C}^{(t)}$ is the centrality matrix for each time-slice $t$ (for the temporal eigenvector centrality, $\textbf{C}^{(t)}$ is the adjacency matrix for time-slice $t$ including all $N$ nodes in the original network), $\textbf{I}$ is the identity matrix of dimension $N \times N$, and $\varepsilon \in (0,\infty)$ dictates how each time-slice is connected to its subsequent one according to the correlation between time-slices. The parameter $\varepsilon \to 0^+$ leads to strongly interconnected time-slices, while $\varepsilon \to \infty$ leads to independent time-slices. 

\bigskip

Since choosing the best value of $\varepsilon$ is non-trivial, and is case-dependent, in our work we set $\varepsilon = 1$ , which essentially sets the weight of self-links between time-slices to 1. Note that a node in time-slice $t$ can either propagate the information forward to $t+1$ ($\textbf{I}$ in the superdiagonal of the matrix $\mathbb{C}(\varepsilon)$) or borrow information from itself in the previous time step $t-1$ ($\textbf{I}$ in the subdiagonal of the matrix $\mathbb{C}(\varepsilon)$). This is important for the correct functioning of eigenvector-based centrality algorithms, as causal coupling (allowing nodes to only connect with themselves either forward or backward in time) can yield non-irreducible (supra-centrality) matrices, which are problematic for the calculus of eigenvectors~\cite{taylor2017eigenvector}. 

The leading eigenvector of the supra-centrality matrix $\mathbb{C}(\varepsilon)$ gives the \textit{joint centrality score} of each node in each time-slice. This allows us to compute from this matrix, two different types of centrality measures: (1) the Marginal Node Centrality (MNC) --- the summary of nodes' scores across all time-slices; and (2) the Marginal Layer Centrality (MLC) --- the summary of centrality scores for a time layer across all nodes. 

As we mentioned earlier, this supra-centrality matrix is applicable to any eigenvector-based centrality; therefore, we also calculate the temporal PageRank by using the same method. PageRank \cite{brin1998anatomy} is the algorithm that underpins Google's search engine, originally proposed by Page and Brin in 1998 as a method for identifying the most frequently accessed webpages on a given subject. It establishes a ranking based on the importance of each page, ensuring that higher-ranked pages appear first in search results. A temporal adaptation of PageRank is computed by the same eigenvector-based centrality method by setting

\begin{equation*}
    \textbf{C}^{(t)}= p\textbf{A}^{(t)}diag(d^{(t)}_1, . . . , d^{(t)}_N)^{-1} + (1 - p)\textbf{v}\textbf{1}^T,
\label{eq:PR_temp}
\end{equation*}

where $d^{(t)}_v = \sum_u \textbf{A}^{(t)}_{uv}$ is the out-degree of node $v$, the quantity $1 - p \in [0, 1]$ is the damping coefficient, \textbf{1} is a vector of ones, and \textbf{v} is the personalized PageRank vector (which is set to be $\textbf{v} = N^{-1}\textbf{1}$). The parameter $p$ is set to $0.85$ as in the original paper~\cite{brin1998anatomy}. Nodes with out-degree 0 are handled by adding a single self-edge for each of these nodes.


Another well-known influence measure is the Katz centrality~\cite{katz1953new}, which we will look into detail next.

\subsubsection{Temporal Katz centrality}

The original paper on the Katz centrality~\cite{katz1953new} calculate people's influence by taking into account not only the number of direct links to each individual but, also, the influence of each individual's neighbours. The method consists of considering all paths of two steps, three steps, and so on, and weighing them to allow for the lower effectiveness of longer chains. Therefore, the impact of a k-step chain is computed by weighing it with $\alpha^k$. In this sense, a k-step chain has probability $\alpha^k$ of being effective, where $\alpha \rightarrow 0$ corresponds to complete attenuation while $\alpha = 1$ corresponds to absence of any attenuation. The influence of nodes in a k-chain network is therefore given by

\begin{equation*}
\alpha A + \alpha^2A^2 +... + \alpha^kA^k = (I - \alpha A)^{-1} - I,
\end{equation*}

which converges to the resolvent matrix $(I -\alpha A)^{-1}$ when $\alpha < 1/ \zeta(A)$~\cite{grindrod2011communicability}. Here $A$ denotes the adjacency matrix, $I$ is the identity matrix of same dimensions as $A$, and $\zeta(A)$ denotes the largest eigenvalue in modulus of the matrix $A$.

Therefore, for simplicity, when calculating Katz centrality we set 

\begin{equation*}
    \alpha = \frac{1}{\zeta(A)} - 10^{-2}, 
\end{equation*}    
    
where $1/\zeta(A)$ is the limiting $\alpha$ value for which Katz centrality is reduced to the eigenvector centrality ~\cite{zhan2017identification}.

Grindrod and Parsons~\cite{grindrod2011communicability} extend the Katz centrality to temporal networks with $t$ time-slices, which is defined as: 

\begin{equation*}
    \mathcal{Q}= (I -  \alpha A^{(1)})^{-1} (I -  \alpha A^{(2)})^{-1} ... (I -  \alpha A^{(t)})^{-1},
\end{equation*} 
where $(I -  \alpha A^{(t)})^{-1}$ is the inverse of the matrix $I -  \alpha A^{(t)}$.

This method deals with large, sparse networks, and allows a message to ``wait'' at a node until a suitable connection appears at a later time~\cite{grindrod2011communicability}. 

The centrality measure that quantifies how effectively a temporal node $n$ can spread information is given by row sums of the matrix $\mathcal{Q}$, i.e.,

\begin{equation*}
    Q_n = \sum_{k = 1}^t \mathcal{Q}_{nk}
\end{equation*}

is the temporal Katz centrality of the temporal node $n$.

\subsubsection{Marginal Community Centrality}

The above discussed centrality measures deal with temporal data, and therefore, to compare influence in polarised communities, we extend the idea of marginal node centrality (MNC) to calculate the marginal community centrality (MCC), i.e., for a community level centrality. The marginal centrality for community $C1$ is computed by aggregating the MNC for each polarised community, as follows:

\begin{equation*}
    MCC(C1) = \frac{\sum \text{MNC of nodes in C1}}{\text{Number of nodes in C1}}.
\label{eq:MCC}
\end{equation*}

Figure~\ref{Fig2}~(b) shows the structure of the table that contains the joint community-time centrality, MLC and MCC measures.

As a benchmark to the centrality methods studied, we use the Temporal Independent Cascade Model, as described next.

\subsection{Temporal Independent Cascade Model}\label{sub:temporal_ICM}

The Susceptible-Infected-Recovered (SIR) model~\cite{hethcote1989three}, or its discrete-time probabilistic equivalent, the Independent Cascade Model~\cite{kempe2003maximizing} (ICM), are commonly used as a benchmark to assess the accuracy of centrality methods~\cite{chen2012identifying, zhao2018identifying, hajarathaiah2023algorithms, fink2023centrality}. In a single Monte Carlo simulation of the ICM, nodes can exist in three states, Susceptible-Infected-Recovered (just as in the case of the SIR model). Every infected node in a discrete time-step has one chance to infect its susceptible network neighbours, with independent probability $\rho \in [0,1]$, before being removed to the recovered state, i.e., a node is in the infected state for only one discrete time-step. If a susceptible node has multiple network neighbours trying to activate it, these attempts occur in a random order. The process terminates once there are no more infected nodes active in a time-step to further propagate the influence. Each node is initially in the susceptible/inactive state. To initialise the process the seed node state is changed to infected/active. Each node is selected as the seed for a large number of Monte Carlo simulations, where the average cascade size is calculated for that seed node. The average cascade size calculated across all nodes is used as a benchmark for the centrality scores, where nodes producing larger cascades, on average, are assumed to be more influential, and as such should have higher centrality scores~\cite{fink2023centrality, nandi2023pew, lu2016vital}. As a benchmark for temporal centrality measures, we use the Temporal Independent Cascade Model (T-ICM) developed by Haldar et al.~\cite{haldar2023temporal}. This is particularly useful as a benchmark for our empirical network, as it allows us to easily calculate benchmark for centralities, where we have a temporally evolving network structure.

The T-ICM introduced by Haldar et al. is a straightforward temporal extension of the classic ICM. In the traditional ICM, infections occur in discrete time-steps: nodes infected at the end of time-step $t$ become the seeds for possible infections at time $t+1$. Haldar et al.~\cite{haldar2023temporal} reformulate this process to account for networks that evolve temporally in discrete time-slices. To model the temporal dynamics, the authors run the ICM on each temporal network, $A^{(t)}$, for one discrete time-step. Any newly infected nodes become the seed infections on the next temporal network, $A^{(t+1)}$, and the process continues until there is no more infected nodes, or the maximum number of times-slices have been reached (i.e., one time-step for each temporal adjacency matrix). This can  can be interpreted as analogous to the matrix $\mathbb{M}$ defined in Eq.~\ref{eq:M_matrix}. Specifically, the T-ICM can be constructed by creating a weighted matrix $\mathbb{W}(\rho)$, where each edge represents the probability of a currently infected node infecting its neighbour. This matrix, $\mathbb{W}(\rho)$, is  a multilayer network, where we have created a separate layer for each time-slice. In this construction, each node at time $t$ is linked to its corresponding node at time $t+1$ (e.g., node $v_i$ in layer $t$ connects to node $v_j$ in layer $t+1$ via an inter-layer link). Additionally, if a node $v$ has an edge to node $u$ in the original graph at time $t$, this is translated as an edge from node $v$ in time-slice $t$ to node $u$ in time-slice $t+1$, resulting in an off-diagonal block structure in the temporal adjacency matrix. Thus, the temporal matrix of infection probabilities $\mathbb{W}(\rho)$ effectively encodes the ICM dynamics across time layers as: 
 
\begin{equation}
    \mathbb{W}(\rho) =
\begin{bmatrix} 
    0 & \textbf{A}^{(1)} \rho + \textbf{I} & 0 & 0 & 0 & ... & 0\\
    0 & 0 & \textbf{A}^{(2)} \rho + \textbf{I} & 0 & 0 & ... & 0\\
    0 & 0 & 0 & \textbf{A}^{(3)} \rho + \textbf{I} & 0 & ... & 0\\
    . & . & . & . & . && . \\
    . & . & . & . & . && . \\
    . & . & . & . & . && . \\
    0 & 0 & 0 & 0 & 0 & ... & \textbf{A}^{(t)} \rho + \textbf{I}\\
    0 & 0 & 0 & 0 & 0 & ... & 0\\
\end{bmatrix},
\label{eq:A_matrix}
\end{equation}
where $\textbf{A}^{(t)}$ is the adjacency matrix of each time-slice $t$, $\textbf{I}$ is the identity matrix of dimension $N \times N$, and $\rho \in (0,1]$ is the probability of an infected node passing the information forward to a neighbour. 

\bigskip

 Applying the ICM on this multilayer network with weighted edges (weights representing infection probabilities) is equivalent to running the ICM separately on each time-slice, for one discrete time-step and using the new infected/activated nodes as the seed nodes for the next time-slices adjacency matrices. It is important to note here that for the purpose of analysing online social networks, we create a slightly modified version of the method developed in~\cite{haldar2023temporal}. In the original method, there are no self-edges of a node between time-slices, i.e., an infected node in time-slice $t$ returns to the susceptible state in $t+1$, and can be reinfected.
 
Matrix $\mathbb{W}(\rho)$ ensures that a node infected in time-slice $t$ will still be infected and will attempt to pass the information forward in the subsequent time-slices. Hence, the identity matrix $I$ added to each weighted block matrix in $\mathbb{W}(\rho)$. This is aligned with the online information spread process, where a content is still available to be seen and spread forward in the future, and cannot spread back to a node previously infected. Infected nodes attempt to infect each neighbour in their own time-slice with probability $\rho$. 

\subsection{Simulation and Empirical Analysis of Polarised Temporal Networks}

Our goal in this paper is to study the dynamics of communities influence in temporal polarised networks by applying a range of centrality measures and using the centrality scores (and the average cascade size via ICM) to aggregate users into bands of influence via clustering techniques. We start our analysis by applying our methods to synthetic networks where we know the true community and influence band structure. This way, we can assess our methods' performance in a controlled setting before applying our techniques to analyse real-world Twitter/X networks, originally studied in~\cite{pena2025finding}.

\subsubsection{BandNet: synthetic polarised network with bands of influence}\label{subsub:bandnet}

To apply a T-ICM and temporal centrality measures to networks with bands of influence in temporally polarised networks, we first create a synthetic network that: (1) Has two communities, where the network contains a small number of cross-community links compared to the number of in-community links, mimicking a polarised environment, as seen in our previous work~\cite{pena2025finding}. (2) Includes users that can be clearly classified into bands of influence, allowing for the comparison of results obtained using various centrality measures in a controlled and simulated setting. This approach enables us to gauge the general behaviour expected when applying different centrality measures to real-world networks. 

Additionally, in real-world networks, a node influence changes over time as edges are created, deleted, or reallocated in time. To simulate this temporal evolution in a network's structure, we follow two steps. In the first step, node influence is changed by swapping nodes between influence bands, effectively swapping the number of connections a node has. In the second step, to capture the creation or deletion of edges, a fraction of the intra-community edges are selected and rewired, and similarly the same fraction of the inter-community edges are rewired. This new  configuration of the network represents a new time-slice. The detailed process is explained below:

\begin{enumerate}
    \item Create the network with communities and bands
    \begin{enumerate}
        \item Create two networks where a small number of nodes with a high degree (band 1), a moderate amount of nodes with a moderate degree (band 2), and a large amount of nodes with a low degree (band 3). Each one of these networks will be a community in the synthetic network to be studied.
        \item Connect these two networks (communities) together by adding a small number of edges between randomly selected nodes in different communities. The number of edges between communities is required to be small compared to the number of edges inside each community, as we are mimicking polarised networks.
    \end{enumerate}
    
    \item Create the temporal evolution
    \begin{enumerate}
        \item To create the temporal evolution of the network, select x\% of nodes uniformly at random from each band and swap their original bands. To better model the behaviour of nodes in a time-evolving network, nodes can only change from one band to its neighbouring band(s), that is, a node originally in band 1 can only change to band 2, a node in band 2 can either change to band 1 or band 3, and a node in band 3 can only change to band 2. In this step, in order to change node $v_1$ from band 1 to band 2, for example, we require a node $v_2$ originally in band 2 to swap places with $v_1$, and become a band 1 node. It is important to note that the number of nodes in each band is maintained.
        \item To make the temporal evolution of the network closer to reality, rewire a percentage of the intra-community edges in the new network time-slice. Repeat for a percentage of the inter-community edges. If by deleting an edge a node becomes disconnected, it is then reconnected with a randomly selected node of its own community.
    \end{enumerate}
\end{enumerate}

\begin{figure}[!ht]
\centering
\includegraphics[width = 0.9\textwidth]{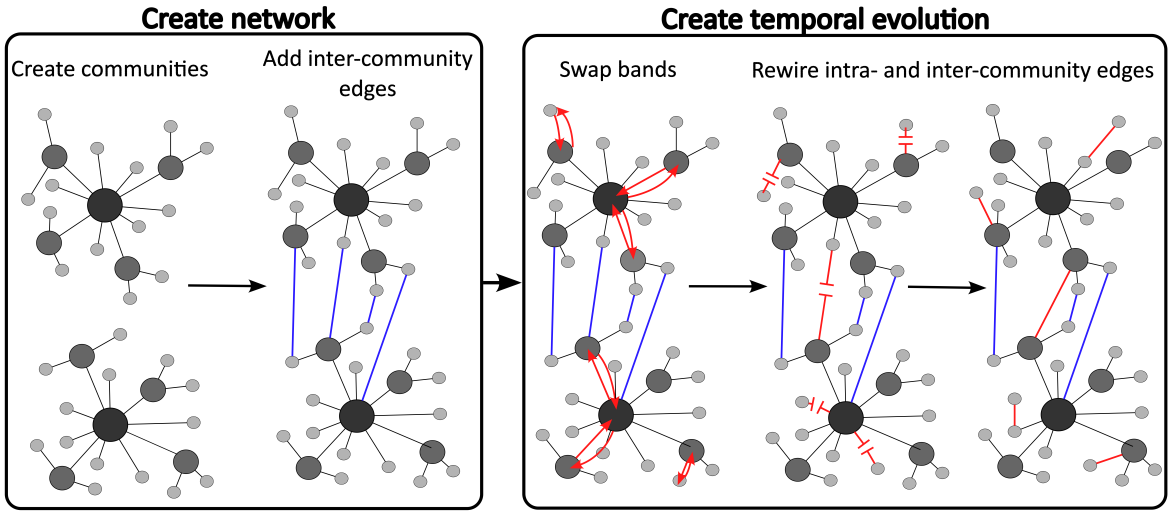}
\caption{{\bf Schematic of the process for building a temporal BandNet.} Here nodes of darker colour and larger size are nodes of greater degree, inter-community edges are blue-coloured, and red edges represent the changes in the network structure.}
\label{Fig3}
\end{figure}

Figure~\ref{Fig3} shows an example figure of this process. In this figure, nodes of darker colour and larger size are nodes of greater degree, inter-community edges are blue-coloured, and red edges represent the changes in the network structure.

After studying examples of the synthetic BandNet networks using the temporal centralities previously discussed, we will analyse a conversation Twitter network, and a randomised version of it, created as explained in \nameref{S1_Appendix}. Random graph models constructed from real networks perform well in estimating quantities investigated, and in some cases give results of high accuracy~\cite{newman2001random}. We will therefore check how centrality measures in a randomised network behave compared to their performance in the original network.

\subsubsection{Properties of studied networks}

We start our analysis with the previously outlined synthetic BandNet network, which contains two communities and well-defined bands of influence for nodes. We will use different degree structures with Bandnet, which will progressively get more complex to address how the centrality methods perform as we increase the complexity. We will start with a fixed set of possible degree each node can take and move to a less homogenous network structure, where each node degree will be sampled from a Poisson distribution, in which we will explore the effect of relative community size on the centrality measures. After which we will also study the RT8 network originally studied in~\cite{pena2025finding} and a random version of this network created by using the configuration model, as described previously. Table~\ref{tab:networks} shows properties of the studied networks.

\begin{table}[!ht]
\centering
\caption{
{\bf Studied networks}}
\begin{tabularx}{\textwidth}{|X|X|X|X|X|X|X|}
\hline
\bf{Network} & \bf{Band 1} & \bf{Band 2} & \bf{Band 3} & \bf{C1} & \bf{C2} & \bf{Inter edges}\\ 
\hline
BandNet1 & $N=10$, $d = 30$ & $N=100$, $d=10$ & $N=1\,000$, $d= 2$ & $N=555$ & $N=555$ & $|E|=100$\\ 
\hline
BandNet2 & $N=10$, $\lambda = 40$ & $N=100$, $\lambda = 20$ & $N=1\,000$, $\lambda = 5$ & $N = 555$ & $N = 555$ & $|E| = 100$\\ 
\hline
BandNet3 & $N=15$, $\lambda = 40$ & $N=150$, $\lambda = 20$ & $N=1\,500$, $\lambda = 5$ & $N = 555$ & $N = 1\,110$ & $|E| = 100$\\ 
\hline
RT8: config. & - & - & - & $N=2\,948$ & $N=463$ & $|E|=7\,353$\\ 
\hline
RT8: original & - & - & - & $N=2\,948$ & $N=463$ & $|E|=7\,353$\\ 
\hline
\end{tabularx}
\begin{flushleft} Here $N$ is the number of nodes; $d$ is the exact degree of each node, homogeneous to each band; $\lambda$ is the average degree for SBM networks; $|E|$ is the number of edges, C1 is the community 1 and C2 is the community 2 in the network.
\end{flushleft}
\label{tab:networks}
\end{table}

We start with a simple example where we create a network using configuration model with two communities of the same size and evolves over four time-slices (BandNet1). Initially, on time-slice 1, band 1 contains 10 nodes (5 from each community) each of degree 30, band 2 contains 100 nodes (50 from each community) each of degree 10, and band 3 contains 1\,000 nodes (500 from each community) each of degree 2. We link the two communities by drawing edges between 100 random nodes in community 1 and 100 random nodes in community 2, sampled without replacement. The subsequent time steps are created by applying step 2(a) of the network creation process, where 10\% of the nodes in each band may change to its neighbour band --- if the node is originally in band 3, it may change to band 2 given band 2 can still take swaps --- and 10\% of inside-community edges and 10\% of in-between communities edges are rewired following step 2(b). 

To increase the complexity of our synthetic network while still keeping it reasonably simple, BandNet2 is a network with two communities of the same size, where nodes follow one of three possible average degree distributions: band 1 contains 5 nodes in each community whose degree is sampled from a Poisson distribution with mean 40, band 2 contains 10 nodes in each community whose degree is sampled from a Poisson distribution with mean 20, and band 3 contains 500 nodes in each community whose node degree is sampled from a Poisson distribution with mean 5. There are 100 inter-community edges. The time evolution is created the same way as before.

BandNet3 analysis aims to understand how influence measures behave when communities are of different sizes. It contains 555 nodes in community 1 and 1\,110 nodes in community 2, that is, community 2 is twice the size of community 1. Its initial structure is as follows: Band 1 contains 5 nodes of community 1 and 10 nodes of community 2 with degree sampled from a Poisson distribution with average 40, band 2 contains 50 nodes of community 1 and 100 nodes of community 2 with degree sampled from a Poisson distribution with average 20, and band 3 contains 500 nodes of community 1 and 1\,000 nodes of community 2 with degree sampled from a Poisson distribution with average 5. There are 100 inter-community edges. The time evolution is created the same way as before.

The real-world RT8 network was constructed from Twitter/X mentions around the Irish Abortion Referendum of 2018, using mentions among the most active users that tweeted using at least one of the tracked hashtags \textit{\#repealthe8th}, \textit{\#savethe8th}, \textit{\#loveboth}, \textit{\#together4yes}, and \textit{\#retainthe8th} from the $1^\textrm{st}$ of May to the $27^\textrm{th}$ of May 2018 (two days after the referendum). In previous analysis~\cite{pena2025finding} polarised communities that represent the yes- and no-vote supporters were found. The network (available in~\cite{Pena_Twitter_data_on_2024}) contains $N_1 = 2\,948$ nodes in community 1 and $N_2 = 463$ nodes in community 2. There are 7\,353 inter-community edges, against 127\,242 edges in-community 1 and 21\,197 edges in-community 2. The same applies to the randomised RT8 network.

\subsection{Classification of nodes into bands of influence}\label{sub:clustering}

After applying the centrality measures on previously described networks, we need to classify the nodes into bands of influence according to their centrality score, for each centrality measure, in order to compare them, with T-ICM being used as the true in the empirical networks. We use a clustering technique to identify groups of nodes that are closely related according to their centrality scores. We apply hierarchical clustering with complete linkage as we are seeking maximal intercluster dissimilarity~\cite{james2023introduction}, i.e., groups that are as further apart from each other as possible to avoid overlaps. Here the clusters correspond to the bands of influence. We then check the optimum number of bands by using the elbow method. As we divide the nodes into three bands throughout our analysis (band 1 consisting of high influential nodes, band 2 consisting of mid-influential nodes, and band 3 consisting of low influential nodes), if the optimum number of bands found through hierarchical clustering is greater than 3, we merge bands together according to their average centrality score until we get 3 bands. In the rare case where the optimum number of clusters is less than 3, we select the cut point equal to 3.

With nodes clustered into bands, we assess the performance of each influence measure according to 1) the true bands for synthetic networks, or 2) the bands classification according to every other influence method for the RT8 network (please note that when we lack ground true we rely on the T-ICM as the benchmark for the other methods). To do so, we use balanced accuracy (BA), a metric used to evaluate the performance of a classification model. It is calculated as the average of correct classifications throughout all classes (or bands, in our study), i.e., 
\begin{equation*}
    BA = \frac{(b_1+b_2+b_3)}{3},
\end{equation*}
where $b_n$ is the number of correctly classified nodes into band $n$. 

In the next section, we show how our proposed method works for the synthetic networks we outlined earlier. Following this, we examine the Twitter conversation network about the Irish Abortion Referendum of 2018 to identify possible bands of influence in the network originally studied in~\cite{pena2025finding}.

\section{Results and discussion}

We now present and discuss the results for three synthetic networks generated by using the method previously explained, and for the RT8 Twitter network as previously summarised.

\subsection{BandNet: synthetic networks}

Our synthetic networks allow us to compare different centrality methods results in a setting where we know the  bands and communities structure. Thanks to the synthetic networks construction, we are able to assess the methods accuracy against true bands, that is, how effective each centrality method, together with the clustering technique, is in capturing which nodes fall into each band over time.

\subsubsection{BandNet1 with communities of the same size and fixed degree values} \label{subsub:BandNet_simple}

\begin{figure}[!ht]
\includegraphics[width = \textwidth]{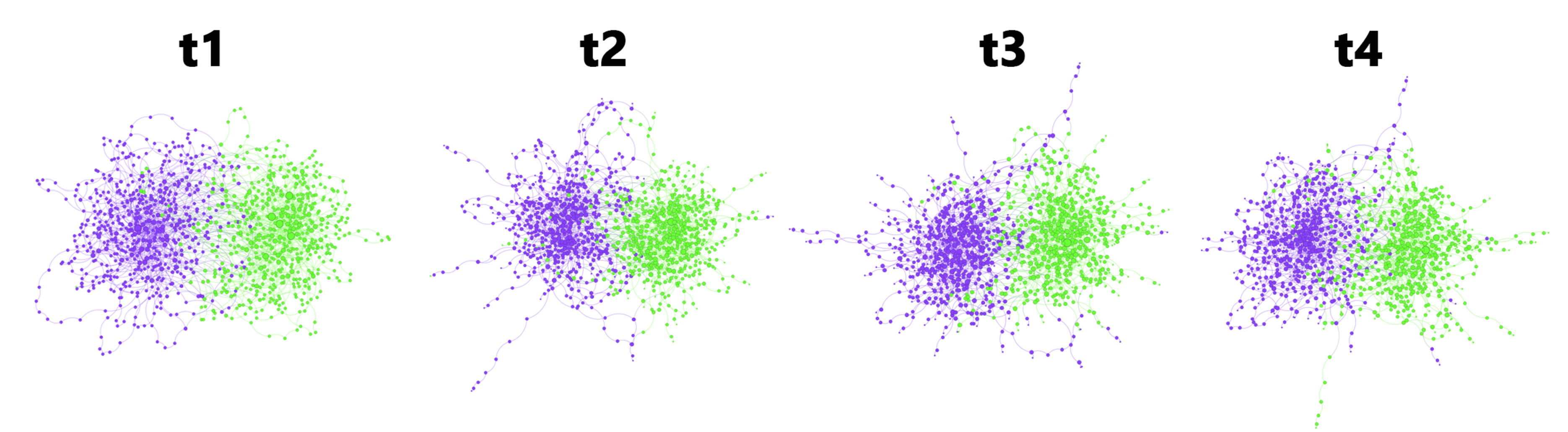}
\caption{{\bf Time-slices of BandNet1.} Nodes are coloured according to the community they belong to, and the size of the node reflects its degree. Layout produced using the Force Atlas algorithm.}
\label{Fig4}
\end{figure}

As mentioned above, we start with BandNet1, a simple example in where we create a network using a configuration model which has two communities of the same size ($N_1 = N_2 = 555$) and evolves over four time-slices. Figure~\ref{Fig4} shows the evolution of the network structure. 

\begin{figure}[!ht]
\centering
\includegraphics[height = 0.75\textheight]{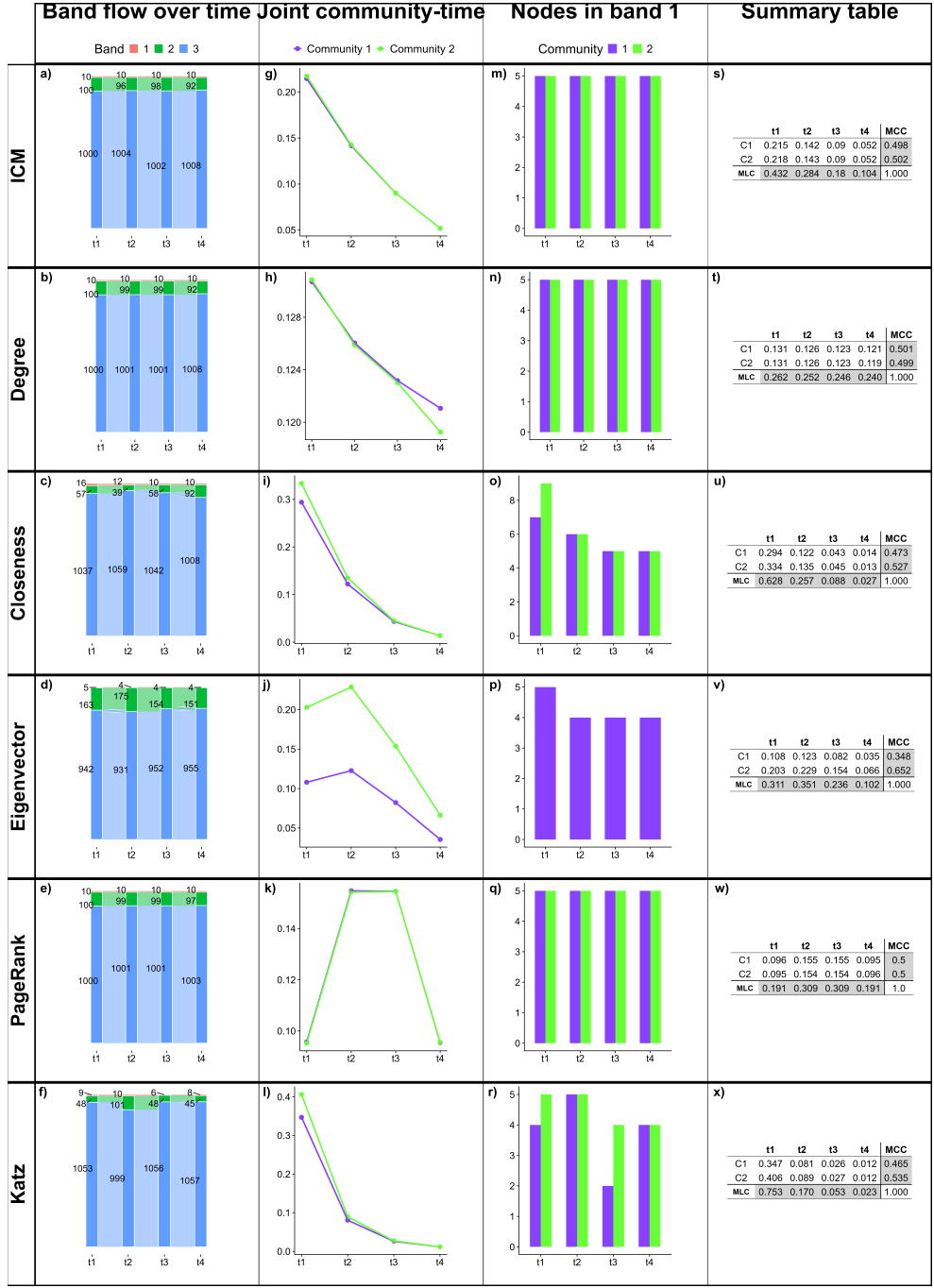}
\caption{{\bf Results for BandNet1.} (a)-(f) how many nodes are in each band in each time-slice, and how nodes move between bands in subsequent time-slices; (g)-(l) normalized influence score for each community over time; (m)-(r) how many nodes of each community are classified in band 1 over time; (s)-(x) summary tables containing the joint community-time scores over time, the MLC over time and the MCC for each community. Here, the infection probability for T-ICM is $\rho = 0.1$ to ensure the process is subcritical.}
\label{Fig5}
\end{figure}

We assess how each centrality measure captures the temporal dynamics of the network by looking at the band flow dynamics over time (Fig.~\ref{Fig5}~(a)-(f)), the joint community-time centrality (Fig.~\ref{Fig5}~(g)-(l)), the nodes in band 1 over time (Fig.~\ref{Fig5}~(m)-(r)), and the summary table containing the joint community-time centrality scores, the MLC and the MCC for each community (Fig.~\ref{Fig5}~(s)-(x)). 

As in BandNet1 nodes can only assume one of three possible degree values, we expect that the centrality methods combined with our band clustering method should be able to capture true bands since nodes swap places with one another, without changing the network structure. The only change in the network structure comes from the rewiring of 10\% of intra-community edges and 10\% of inter-community edges between time-slices. Comparing results for the band flow over time (Fig.~\ref{Fig5}~(a)-(f)), we see that T-ICM, degree centrality and PageRank capture this behaviour well when keeping bands of similar sizes over time. Closeness, eigenvector and Katz centralities also perform well in capturing this temporal dynamics, but with less accuracy. We therefore show that we can successfully aggregate nodes into bands of influence (research question 1). 

To answer research question 2 (``Does the overall influence of a specific community within a polarised network change over time?''), from the joint community-time scores (Fig.~\ref{Fig5}~(g)-(l)) and tables on the fourth column, we conclude that (1) T-ICM, degree, closeness and Katz present similar behaviour, with scores decaying over time. This is partially explained by the fact that in these methods a piece of information starting in time-slice 1 has the chance to spread until time-slice 4, whereas a piece of information that starts in time-slice 4 can only spread through its own time-slice, as it is the last time-slice. The eigenvector-based centralities (Eigenvector and PageRank), on the other hand, consider not only the subsequent time-slices but also the previous ones and tend to assign higher scores to the central time-slices~\cite{taylor2017eigenvector}; (2) Eigenvector centrality consistently gives significantly higher scores to nodes in community 1, which is an indication that it does not perform well in networks with communities. This is supported by the fact that eigenvector centrality can be used as community detection in networks with high enough modularity~\cite{sharkey2019localization, ditsworth2019community}. 

The third column of Fig.~\ref{Fig5} helps us answer research question 3 (``Can we determine which polarised community the most influential nodes belong to, and how do influential nodes differ across communities?''). T-ICM, degree centrality and PageRank show the same number of nodes (5 nodes) from each community in band 1, which remains the same over time. This is the expected result as communities are of same size. The summary tables in the fourth column of Fig.~\ref{Fig5} show that the marginal community centrality (MCC) is similar (close to 50\%) for both communities in all centrality methods except eigenvector. This is expected as communities are of the same size and bands should remain the same (or similar) size throughout the temporal dynamics. Eigenvector centrality returns different MCC values for each community as it is not the most appropriate method for networks with communities as previously pointed out.

Table~\ref{tab:acc_BandNet1} shows the balanced accuracy for the investigated methods against the true bands in BandNet1. Here the true bands are tracked over time from the initial setup in time-slice $t_1$, i.e., nodes that swap bands are tracked over time. T-ICM, degree and PageRank are the methods which score the highest against true bands with an overall balanced accuracy of $0.9$. Closeness and Katz follow closely, and eigenvector centrality scores much lower (overall $0.67$). Time-slice $t_1$ has the highest balanced accuracy for every method, except closeness. This is an expected behaviour as rewiring hasn't occurred at the initial setup of $t_1$, and bands are more clearly laid-out.

\begin{table}[!ht]
\centering
\caption{
{\bf Balanced accuracy in BandNet1}}
\begin{tabular}{|l|r|r|r|r|r|}
\hline
\bf{Method} & \bf{t1} & \bf{t2} & \bf{t3} & \bf{t4} & \bf{Overall}\\
\hline
T-ICM & 1.00 & 0.92 & 0.85 & 0.82 & 0.90\\
\hline
Degree & 1.00 & 0.93 & 0.86 & 0.82 & 0.90\\
\hline
Closeness & 0.83 & 0.85 & 0.85 & 0.82 & 0.84\\
\hline
Eigenvector & 0.80 & 0.70 & 0.60 & 0.59 & 0.67\\
\hline
PageRank & 1.00 & 0.93 & 0.86 & 0.82 & 0.90\\
\hline
Katz & 0.96 & 0.79 & 0.79 & 0.79 & 0.83\\
\hline
\end{tabular}
\begin{flushleft} Balanced accuracy of centrality methods when compared to simulated true bands in BandNet1. The overall balanced accuracy for each method is computed as the average balanced accuracy of time-slices $[t_1,t_4]$.
\end{flushleft}
\label{tab:acc_BandNet1}
\end{table}

\subsubsection{BandNet2 with communities of the same size and Poisson degree distribution}\label{subsub:SBM_samesize}

\begin{figure}[!ht]
\includegraphics[width = \textwidth]{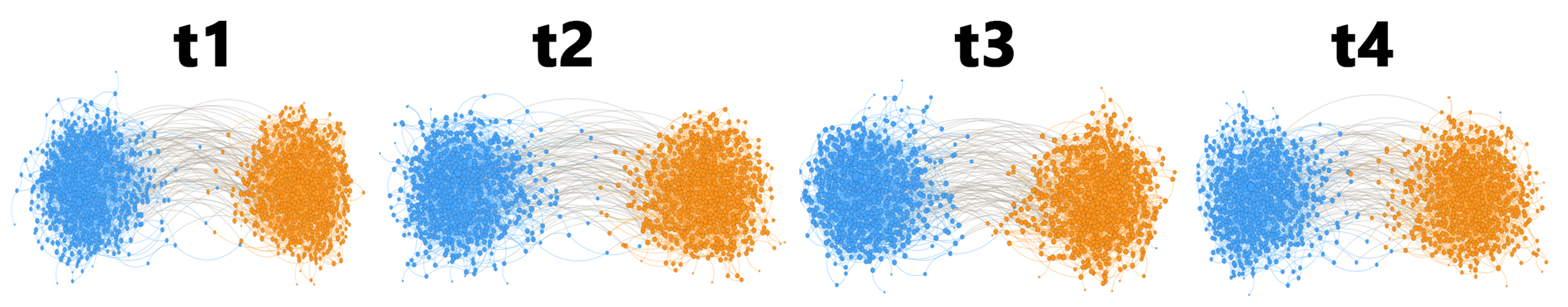}
\caption{{\bf Time-slices of BandNet2.} Nodes are coloured according to the community they belong to, and the size of the node reflects its degree. Layout produced using the Force Atlas algorithm.}
\label{Fig6}
\end{figure}

As a natural and simple extension to our synthetics networks, BandNet2 has two communities of the same size ($N_1 = N_2 = 555$), where the nodes degrees in each band are drawn from a Poisson distribution, as explained previously. Figure~\ref{Fig6} illustrates the network time-slices.

\begin{figure}[!ht]
\centering
\includegraphics[width = 0.6\textwidth]{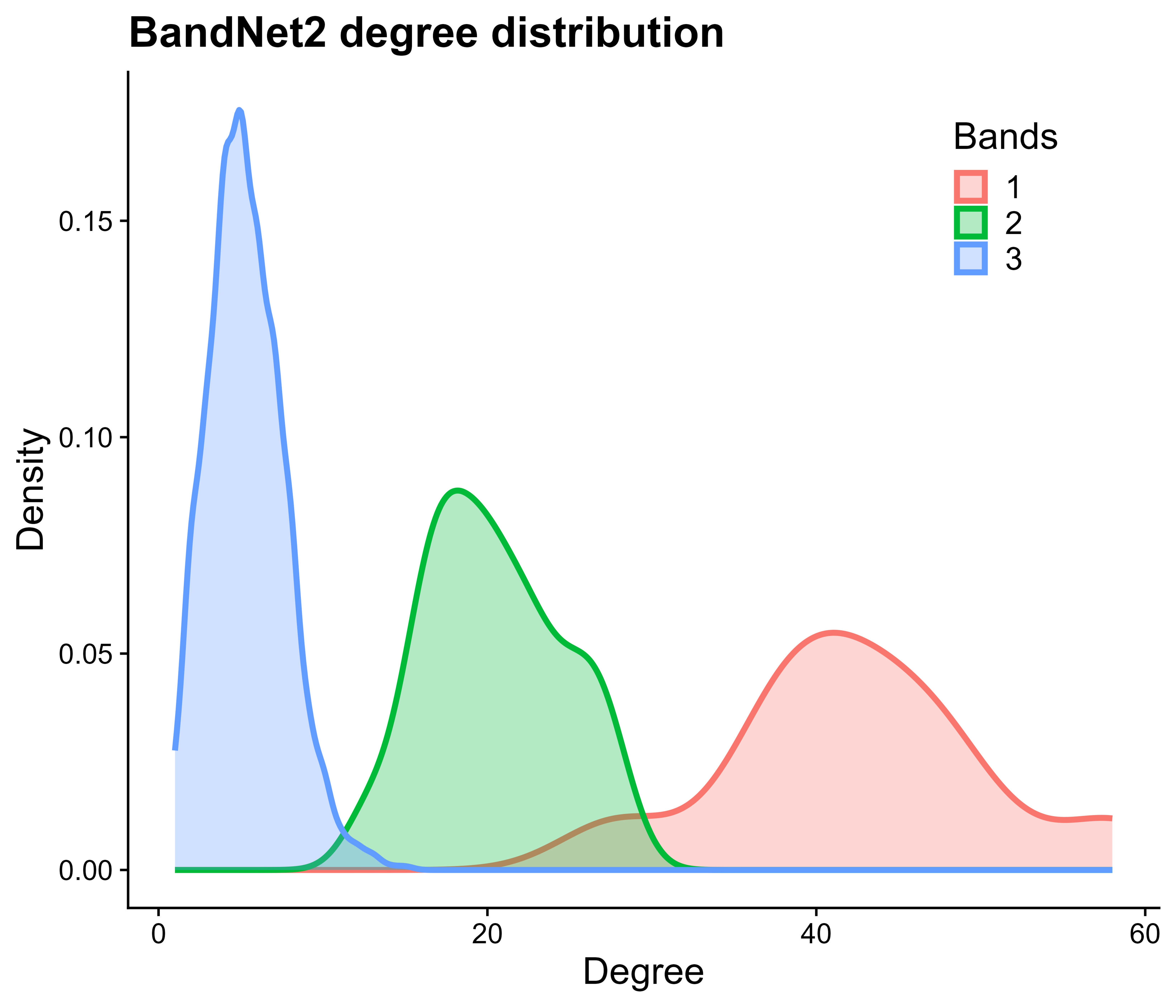}
\caption{{\bf Degree distribution of bands in the initial set up of BandNet2.} Here, band 1 has average degree of $\lambda = 40$, band 2 has average degree of $\lambda = 20$, and band 3 has average degree of $\lambda = 5$.}
\label{Fig7}
\end{figure}

\begin{figure}[!ht]
\centering
\includegraphics[height = 0.75\textheight]{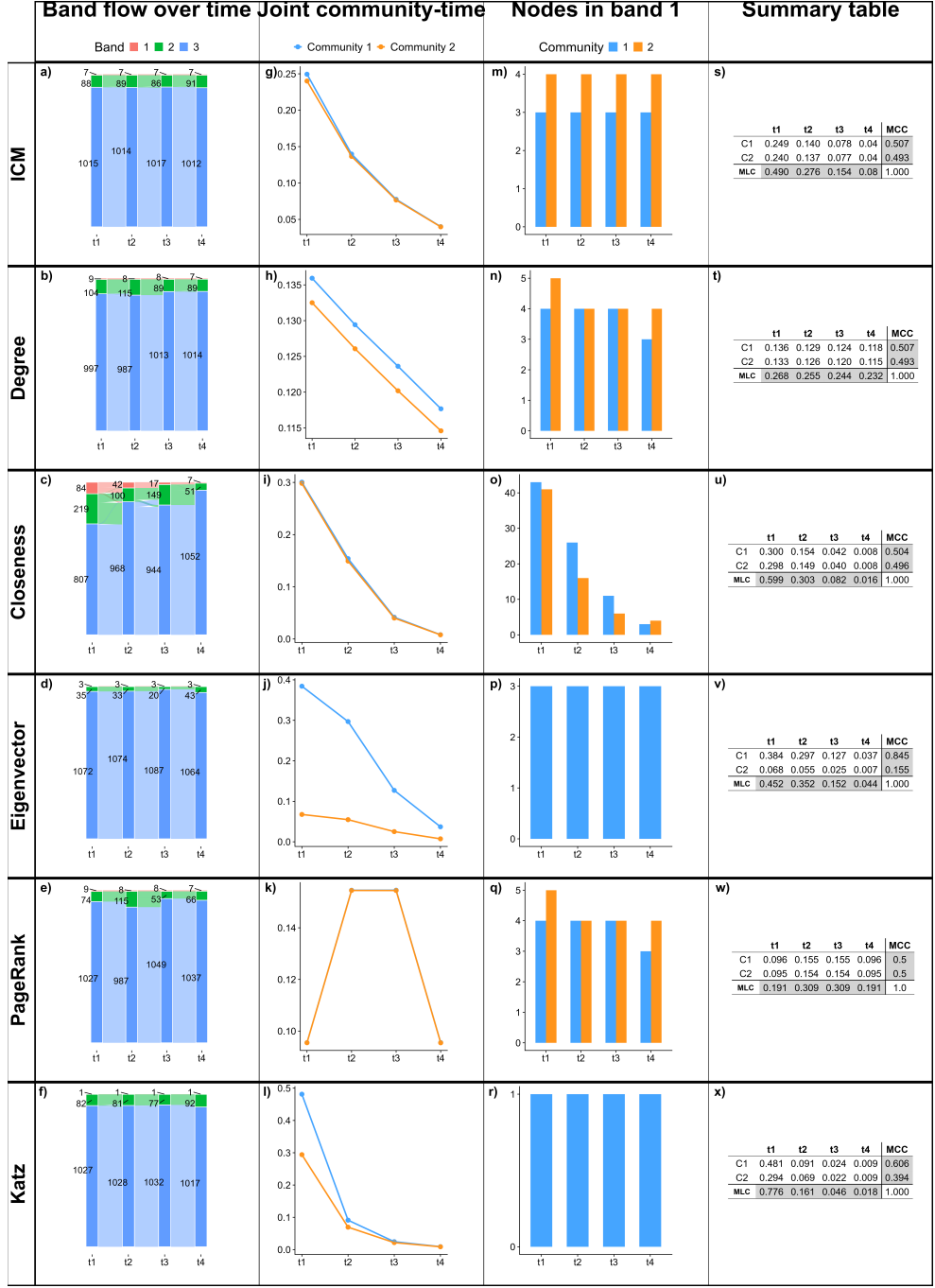}
\caption{{\bf Results for BandNet2.} (a)-(f) how many nodes are in each band in each time-slice, and how nodes move between bands in subsequent time-slices; (g)-(l) normalized influence score for each community over time; (m)-(r) how many nodes of each community are classified in band 1 over time; (s)-(x) summary tables containing the joint community-time scores over time, the MLC over time and the MCC for each community. Here, the infection probability for T-ICM is $\rho = 0.08$ to ensure the process is subcritical.}
\label{Fig8}
\end{figure}

The results of T-ICM and temporal centrality methods on BandNet2 are shown in Fig.~\ref{Fig8}. As the bands degree distributions overlap each other (Fig.~\ref{Fig7}), we expect more variability in the results compared to the BandNet1 results. In fact, although T-ICM and degree centrality (Fig.~\ref{Fig8}~(a) and (b)) still show bands of consistent sizes over time, the initial setup of 10 nodes in band 1, 100 nodes in band 2 and 1\,000 nodes in band 3 is not perfectly captured. PageRank, which was very successful in capturing the initial setup and keep bands of the same size over time in BandNet1, now does not capture well the initial setup and shows bands that fluctuate more in size over time (Fig.~\ref{Fig8}~(e)). Katz centrality (Fig.~\ref{Fig8}~(f)) is successful in maintaining bands of similar sizes throughout the temporal network, however it does not capture the initial setup, and band 1 consists of only one user. Closeness and eigenvector centralities, on the other hand, show bands that vary greatly in size over time (Fig.~\ref{Fig8}~(c) and (d)).

The behaviour of the joint community-time centrality scores (Fig.~\ref{Fig8}~(g)-(l) and (s)-(x)) is similar to the one observed in BandNet1, with T-ICM, degree, closeness and Katz centralities showing a descendent behaviour over time, PageRank giving higher scores to the mid-time-slices, and eigenvector centrality consistently attributing higher scores to nodes in community 1. As per the nodes in band 1 (Fig.~\ref{Fig8}~(m)-(r)), T-ICM is successful in capturing a consistent amount of nodes in each community over time. Degree centrality and PageRank also capture this dynamics well, with small deviations in $t_1$ and $t_4$. Closeness centrality, however, shows a downward trend on the number of nodes in band 1 overall, in both communities. Eigenvector centrality, similarly to what was observed in BandNet1, attributes the highest scores to nodes in community 1, therefore only nodes in C1 are present in band 1. Katz centrality also shows only community 1 in band 1, however this is due to its band 1 having one node only. MCC scores (Fig.~\ref{Fig8}~(s)-(x)) tell us that communities have the exact same average influence over time according to PageRank, and very similar influence according to T-ICM, degree and closeness centralities. Eigenvector and Katz, on the other hand, attribute higher influence to community 1, i.e., MCC is higher for C1 when compared to C2.

According to Table~\ref{tab:acc_BandNet2}, PageRank, T-ICM and degree centrality score the highest overall balanced accuracy ($\approx 0.85$), when compared to the tracked true bands. Eigenvector, Katz and closeness centralities score lower, in this order. Eigenvector centrality scores higher in this network when compared to BandNet1, while all other methods score slightly lower, which is due to the overlapping of degree distributions (Fig.~\ref{Fig7}), resulting in the higher variability in the degree structure for this network, as previously pointed out.

\begin{table}[!ht]
\centering
\caption{
{\bf Balanced accuracy in BandNet2}}
\begin{tabular}{|l|r|r|r|r|r|}
\hline
\bf{Method} & \bf{t1} & \bf{t2} & \bf{t3} & \bf{t4} & \bf{Overall}\\
\hline
T-ICM & 0.98 & 0.89 & 0.80 & 0.72 & 0.85\\
\hline
Degree & 0.98 & 0.87 & 0.82 & 0.72 & 0.85\\
\hline
Closeness & 0.44 & 0.58 & 0.62 & 0.75 & 0.60\\
\hline
Eigenvector & 0.96 & 0.79 & 0.76 & 0.64 & 0.79\\
\hline
PageRank & 0.99 & 0.87 & 0.83 & 0.73 & 0.86\\
\hline
Katz & 0.92 & 0.57 & 0.53 & 0.53 & 0.64\\
\hline
\end{tabular}
\begin{flushleft} Balanced accuracy of centrality methods when compared to simulated true bands in BandNet2. The overall balanced accuracy for each method is computed as the average balanced accuracy of time-slices $[t_1,t_4]$.
\end{flushleft}
\label{tab:acc_BandNet2}
\end{table}

We now analyse results for a network with Poisson distributions and communities of different sizes to understand the impact of community size on the influence in the network as a whole.

\subsubsection{BandNet3 with communities of different sizes and Poisson degree distribution}\label{subsub:SBM_diffsize}

\begin{figure}[!ht]
\includegraphics[width = \textwidth]{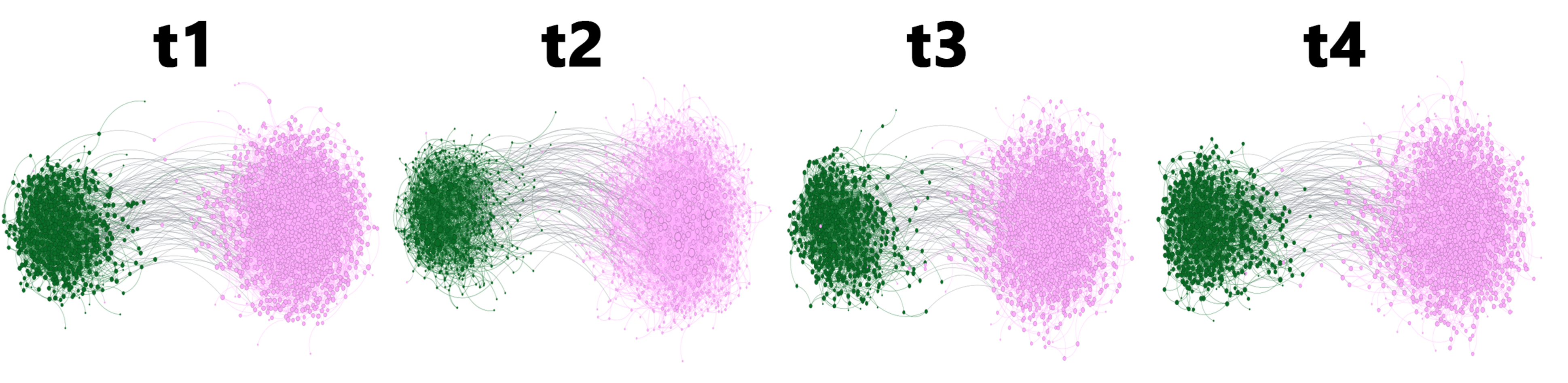}
\caption{{\bf Time-slices of BandNet3.} Nodes are coloured according to the community they belong to, and the size of the node reflects its degree. Layout produced using the Force Atlas algorithm.}
\label{Fig9}
\end{figure}

BandNet3 consists of a network with two communities, where C1 is half the size of C2, i.e., $N_1 = 555$ nodes and $N_2=1\,110$ nodes. The degree of the nodes is Poisson distributed as described in Table~\ref{tab:networks}. In the initial configuration $t_1$, band 1 has 5 nodes in C1 and 10 nodes in C2, band 2 has 50 nodes in C1 and 100 nodes in C2, and band 3 has 500 nodes in C1 and 1\,000 nodes in C2. Figure~\ref{Fig9} illustrates the network over 4 time-slices.

\begin{figure}[!ht]
\centering
\includegraphics[height = 0.75\textheight]{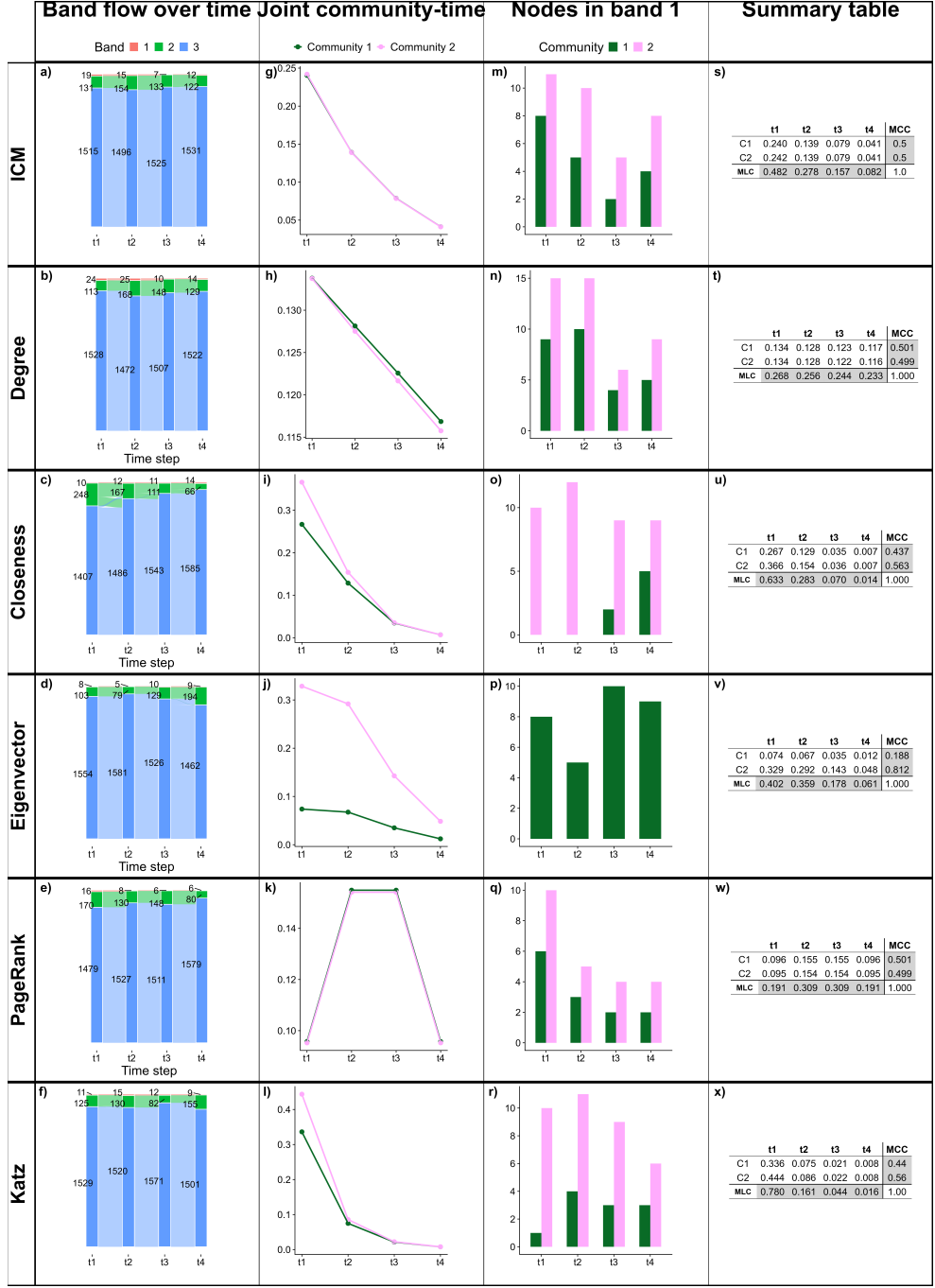}
\caption{{\bf Results for BandNet3.} (a)-(f) how many nodes are in each band in each time-slice, and how nodes move between bands in subsequent time-slices; (g)-(l) normalized influence score for each community over time; (m)-(r) how many nodes of each community are classified in band 1 over time; (s)-(x) summary tables containing the joint community-time scores over time, the MLC over time and the MCC for each community. Here, the infection probability for T-ICM is $\rho = 0.08$ to ensure the process is subcritical.}
\label{Fig10}
\end{figure}

Figure~\ref{Fig10} shows the results of the T-ICM and temporal centrality methods on BandNet3. Similarly to BandNet2, T-ICM and degree centrality  show bands of reasonable consistent sizes over time (Fig.~\ref{Fig10}~(a) and (b)), however $t_1$ is not well captured, and it presents higher deviations when compared to BandNet2. PageRank shows bands that fluctuate more in size over time than for the previous examples of BandNet (Fig.~\ref{Fig10}~(e)). Katz centrality (Fig.~\ref{Fig10}~(f)) captures better the initial 15-150-1\,500 band sizes setup when compared to its performance in BandNet2; however, band 2 in $t_3$ is considerably smaller than in the other time-slices. Closeness and eigenvector centralities show bands that vary greatly in size over time and are not very successful in capturing the $t_1$ configuration (Fig.~\ref{Fig10}~(c) and (d)).

The behaviour of the joint community-time centrality scores (Fig.~\ref{Fig10}~(g)-(l) and (s)-(x)) is similar to the ones observed in BandNet1 and BandNet2, with T-ICM, degree, closeness and Katz centralities showing a descendent behaviour over time, PageRank giving higher scores to the central time-slices, and eigenvector centrality consistently attributing higher scores to nodes in one of the communities, this time the larger community C2. 

The results are quantitatively different for the nodes in band 1 (Fig.~\ref{Fig10}~(m)-(r)), where every method place a higher number of nodes in the largest community C2 in band 1, except eigenvector centrality. According to the eigenvector centrality, band 1 is entirely composed by nodes in the smallest community C1. This is due toThis causes the localization of eigenvector
centrality commonly seen in modular networks the information getting confined through random walks in C1, given its high modularity~\cite{sharkey2019localization}, or due to many nodes potentially scoring zero if they have no inward edge or only inward edges coming from nodes with zero scores~\cite{newman2018networks}, when calculating eigenvectors in directed networks. Closeness centrality, however, attributes only nodes in C2 to band 1 in the first time-slices $t_1$ and $t_2$. MCC scores (Fig.~\ref{Fig10}~(s)-(x)) show that communities have the exact same average influence over time according to T-ICM or very similar influence according to degree and PageRank centralities. This may be explained by the fact that bands are of proportionate sizes in both communities, i.e., each band in C1 is initially set up to be half the size of the bands in C2, the same proportion $N_1/N_2$ of the number of nodes between communities. Closeness and Katz centralities attribute slightly higher scores to the largest community C2, while eigenvector attributes a much higher influence score to C2, despite band 1 being composed by nodes in C1 only, i.e., the largest number of nodes in C2 biases the scores overall.

When compared to the true bands, Table~\ref{tab:acc_BandNet3} shows that PageRank scores the highest balanced accuracy ($0.88$) among the methods, followed by T-ICM and Katz centrality, which score $0.81$, degree centrality with $0.78$, eigenvector centrality with $0.77$ and lastly closeness centrality with $0.73$. 

\begin{table}[!ht]
\centering
\caption{
{\bf Balanced accuracy in BandNet3}}
\begin{tabular}{|l|r|r|r|r|r|}
\hline
\bf{Method} & \bf{t1} & \bf{t2} & \bf{t3} & \bf{t4} & \bf{Overall}\\
\hline
T-ICM & 0.88 & 0.81 & 0.86 & 0.70 & 0.81\\
\hline
Degree & 0.87 & 0.73 & 0.81 & 0.71 & 0.78\\
\hline
Closeness & 0.63 & 0.73 & 0.84 & 0.70 & 0.73\\
\hline
Eigenvector & 0.85 & 0.85 & 0.75 & 0.64 & 0.77\\
\hline
PageRank & 0.91 & 0.95 & 0.90 & 0.75 & 0.88\\
\hline
Katz & 0.89 & 0.80 & 0.82 & 0.71 & 0.81\\
\hline
\end{tabular}
\begin{flushleft} Balanced accuracy of centrality methods when compared to simulated true bands in BandNet3. The overall balanced accuracy for each method is computed as the average balanced accuracy of time-slices $[t_1,t_4]$.
\end{flushleft}
\label{tab:acc_BandNet3}
\end{table}

From the analysis of our synthetic networks we conclude that neither eigenvector or closeness centralities are appropriate to compute the influence of nodes in a polarised temporal network, and PageRank consistently performs well in this type of network. Next we will analyse the results of T-ICM and the centrality methods here studied in the real RT8 network composed of Twitter mentions on the Irish Abortion Referendum of 2018.

\subsection{Empirical network: the Irish abortion referendum Twitter network}\label{sub:RT8_intro}

\begin{figure}[!ht]
\includegraphics[width = \textwidth]{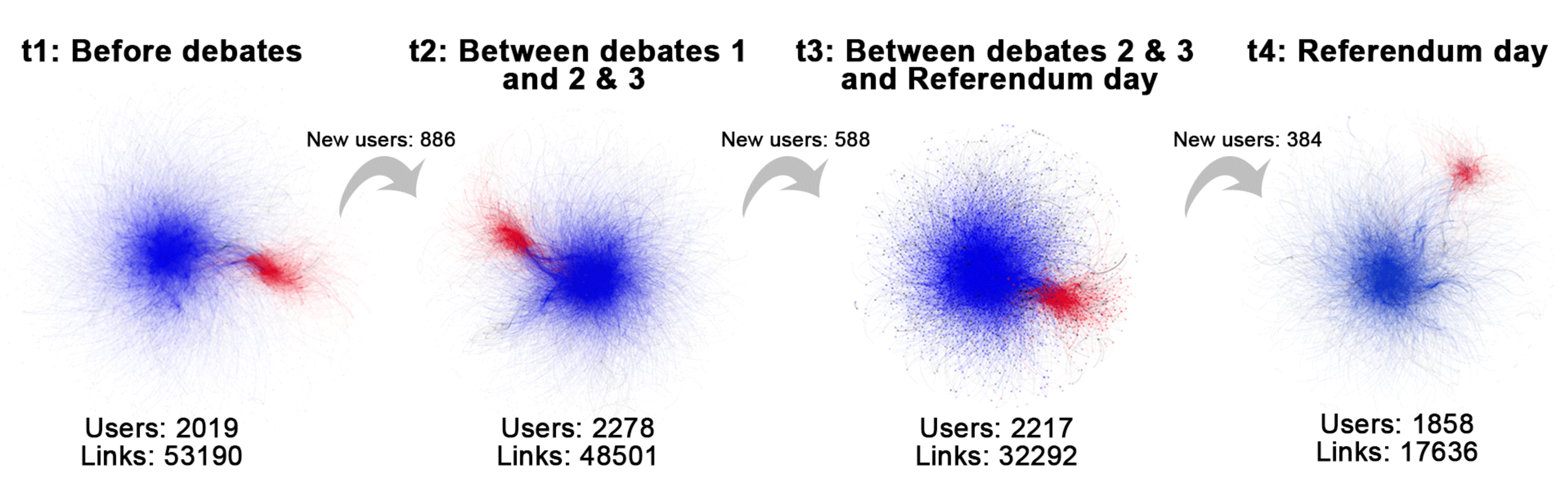}
\caption{{\bf Time-slices on the Irish Abortion Referendum mentions network.} Time-slices of the network showing the number of users, the number of new users coming into the conversation, and the number of links among users in each time-slice.}
\label{Fig11}
\end{figure}

The Twitter mentions network on the Irish Abortion Referendum of 2018 was studied in~\cite{pena2025finding}, where the authors showed a clearly polarised environment. In the context of a referendum, there is a clear community of users that supports the Yes vote, and another clear community that supports the No vote. The network has $3\,411$ nodes --- $2\,948$ in the Yes community and $463$ in the No community, connected by $155\,803$ edges. Holistically, we can consider four time-slices according to important events that affect the network (Fig.~\ref{Fig11}). There were three televised debates, two of them occurred on the same day, therefore the time steps are $t_1$) before debates, $t_2$) after debate 1 and before debates 2 and 3, $t_3$) after debates 2 and 3 and before the referendum day, and $t_4$) on the referendum day.

\subsubsection{The original Irish abortion referendum network} \label{subsub:RT8_original}

\begin{figure}[!ht]
\includegraphics[width = \textwidth]{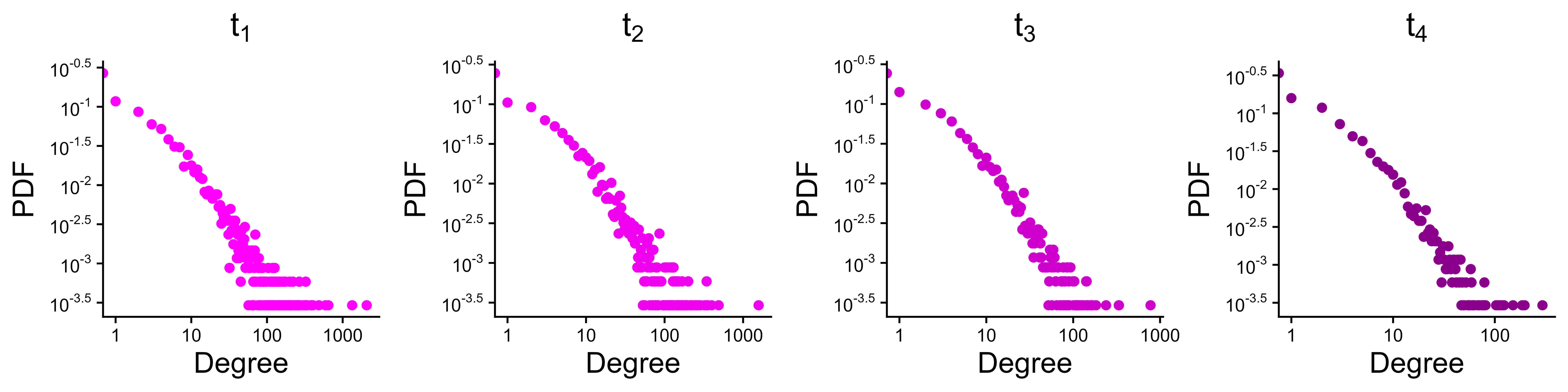}
\caption{{\bf Probability Distribution Function (PDF) of nodes degree in the RT8 original network time-slices.}}
\label{Fig12}
\end{figure}

Real-world networks often display complex structures. Online social networks, in particular, often present heavy-tailed degree distributions~\cite{pena2025finding, o2017integrating, goel2016structural, wu2024power}, where there are many nodes with only a few edges and a few nodes (hubs) with a large number of edges~\cite{posfai2016network}. This type of degree distribution is usually called power-law or scale-free distribution, and looks like the time-slices degree distributions of our RT8 network (Fig~\ref{Fig12}). 

\begin{figure}[!ht]
\centering
\includegraphics[height = 0.75\textheight]{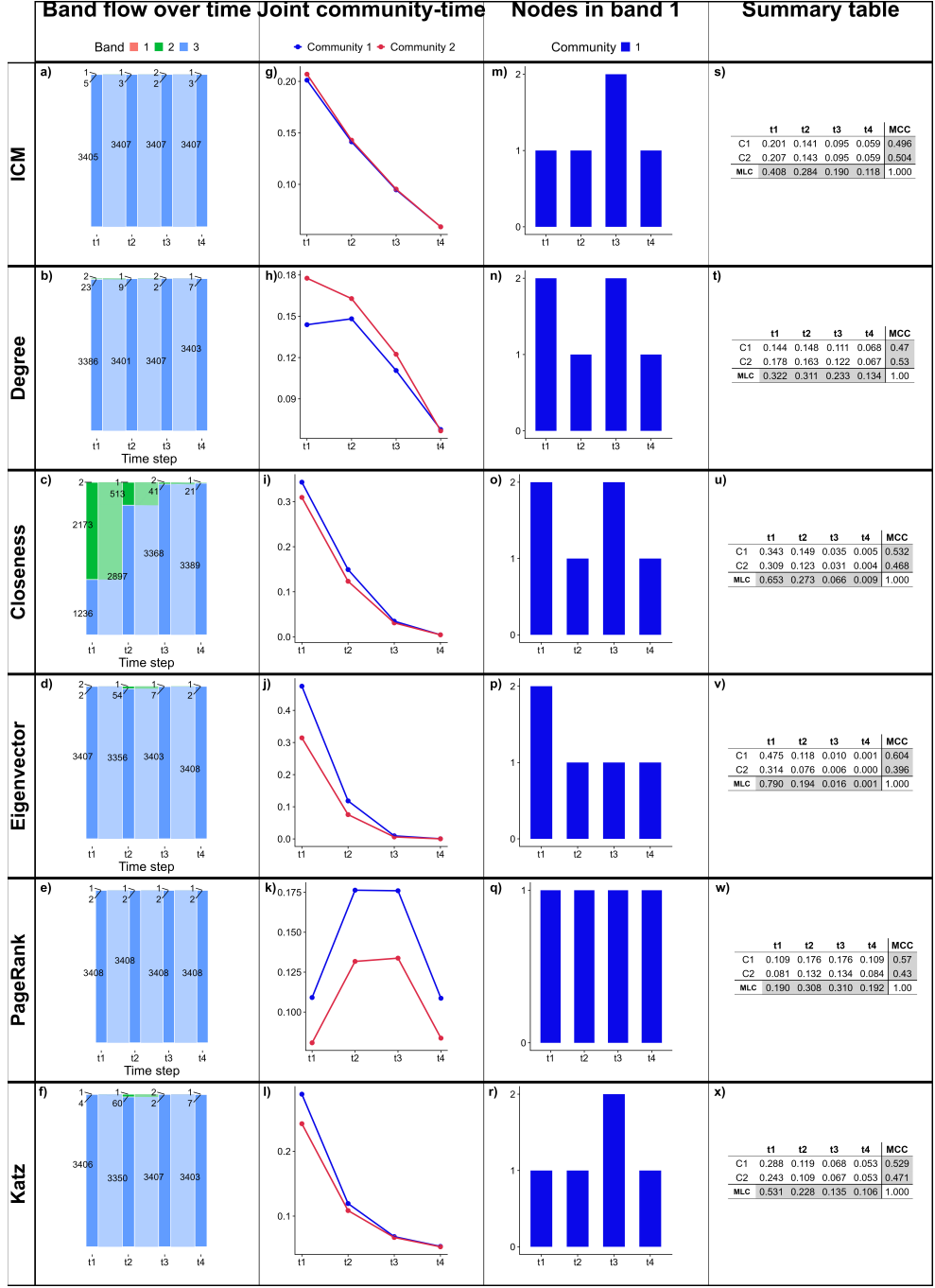}
\caption{{\bf Results for the original RT8 network.} (a)-(f) how many nodes are in each band in each time-slice, and how nodes move between bands in subsequent time-slices; (g)-(l) normalized influence score for each community over time; (m)-(r) how many nodes of each community are classified in band 1 over time; (s)-(x) summary tables containing the joint community-time scores over time, the MLC over time and the MCC for each community. Here, the infection probability for T-ICM is $\rho = 0.02$ to ensure the process is subcritical.}
\label{Fig13}
\end{figure}

In this environment, since only a few nodes present much higher degree distribution than the vast majority of the nodes in the network, we expect that the highest bands, band 1 and 2, are significantly narrower than band 3, which should encompass the vast majority of nodes. Analysing Fig.~\ref{Fig13}~(a)-(f), we see that every method is able to capture this behaviour, except closeness centrality. 

The joint community-time scores shown in Fig.~\ref{Fig13}~(g)-(l) and (s)-(x) follow similar behaviours as for the synthetic networks previously studied, except for eigenvector centrality, that shows a descending behaviour over time (we previously observed the highest scores in $t_2$). Degree centrality attributes slightly higher scores to nodes in C2, which is the opposite behaviour shown by closeness, eigenvector, PageRank and Katz. This is due to C2 presenting tighter connected nodes than C1; therefore, the average degree by community gives higher scores to C2. Closeness, eigenvector, PageRank and Katz centralities attribution of higher scores to node in C1 suggest that these methods are more sensible to the size of communities and tend to give higher scores to the largest community. This translates into the MCC scores, where these methods attribute higher scores to C1 --- the highest difference of scores being in eigenvector centrality --- while degree attributes higher MCC score to C2. T-ICM, however, attributes similar influence over time to both communities, i.e., the communities MCC scores are similar to each other and close to $0.5$.

Every method captured only one or two nodes in band 1 in each time-slice, and these nodes are from C1. Eigenvector captures the same node in band 1 throughout the time-slices, as well as PageRank. However, the nodes capture by each method differ. The node captured by eigenvector is also captured by T-ICM, degree, closeness and Katz up to time-slice $t_3$, and this node presents high out-degree but low in-degree, that is, they mention many users in the network but are rarely mentioned by other users. The node captured by PageRank is a highly active user in canvassing for the Yes vote. They mention and are mentioned by many users in the network and effectively act as a hub of information. This node is not captured in band 1 by any other method apart from PageRank. Other users captured in band 1 are (1) an influential Irish novelist, (2) an active user canvassing for the Yes vote, and (3) a user that is now suspended on X, which may be a suggestion of a bot.

\begin{figure}[!ht]
\includegraphics[width = 0.8\textwidth]{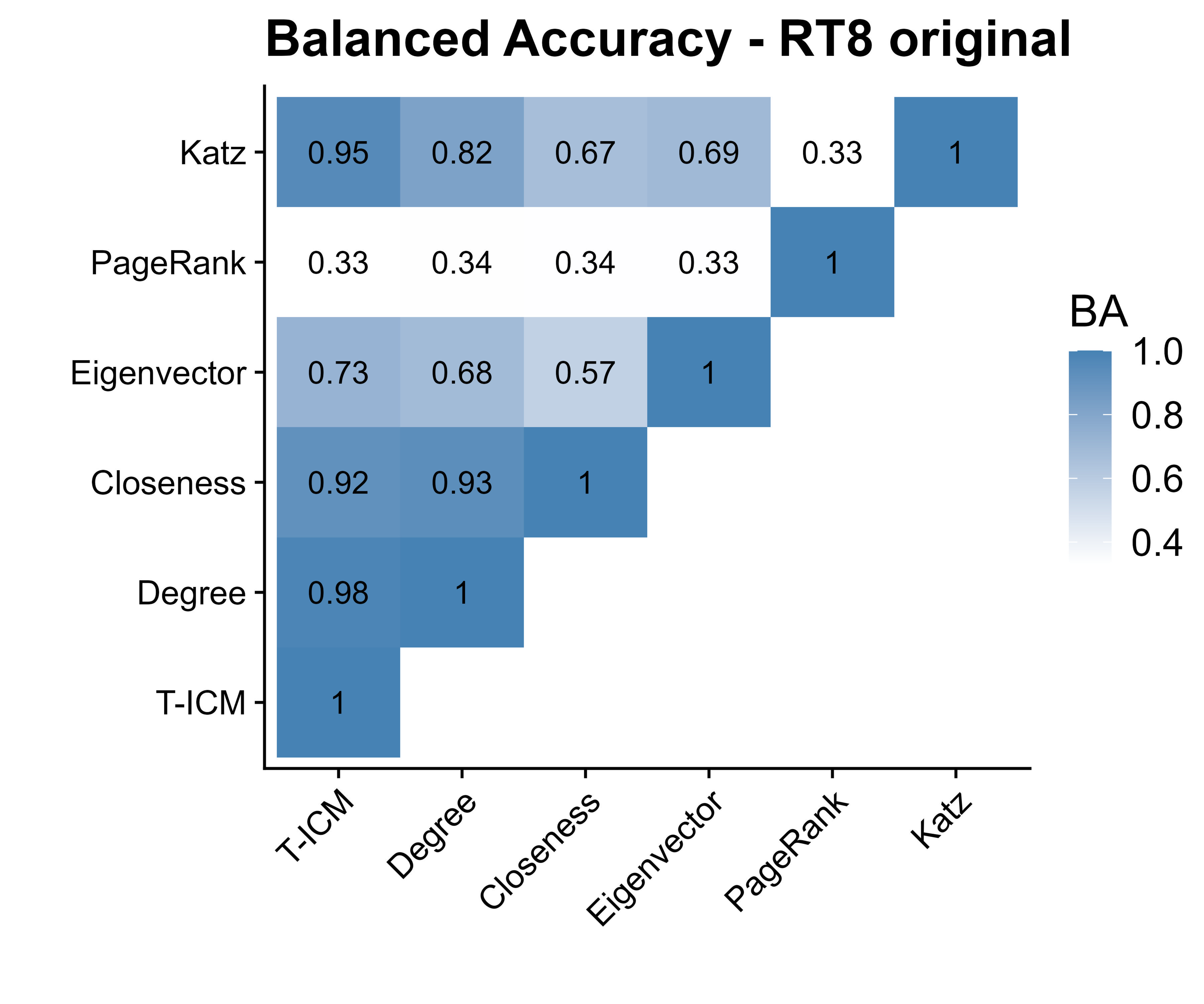}
\caption{{\bf Balanced accuracy between pairs of influence methods in the original RT8 network.} Darker colours represent higher balanced accuracy.}
\label{Fig14}
\end{figure}

Figure~\ref{Fig14} shows the balanced accuracy between pairs of influence methods. PageRank diverges greatly from other methods, while Katz, closeness and degree show good agreement with the benchmark T-ICM.

\subsubsection{Configuration model on the Irish abortion referendum network} \label{subsub:RT8_config}

\begin{figure}[!ht]
\centering
\includegraphics[height = 0.75\textheight]{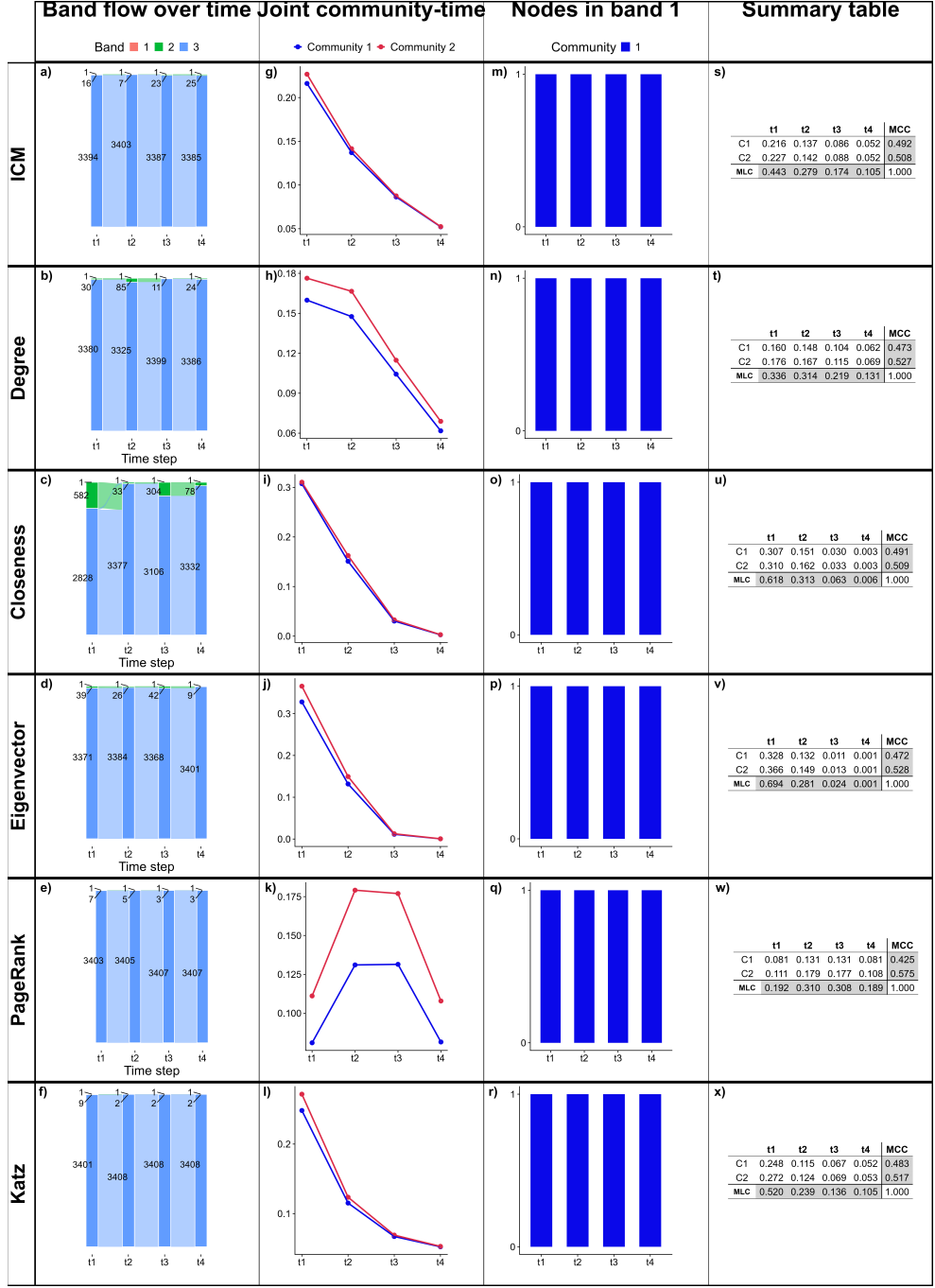}
\caption{{\bf Results for the randomised RT8 network.} (a)-(f) how many nodes are in each band in each time-slice, and how nodes move between bands in subsequent time-slices; (g)-(l) normalized influence score for each community over time; (m)-(r) how many nodes of each community are classified in band 1 over time; (s)-(x) summary tables containing the joint community-time scores over time, the MLC over time and the MCC for each community. Here, the infection probability for T-ICM is $\rho = 0.02$ to ensure the process is subcritical.}
\label{Fig15}
\end{figure}

The configuration model of a network is a way to simplify its structure while maintaining important properties of the network, as previously outlined. Figure~\ref{Fig15} shows the results for T-ICM and centrality measures on the randomised RT8 network, which was built through the use of the configuration model.

As for the original network, bands 1 and 2 are very narrow when compared to band 3, as the overall degree of the network is maintained and is still heavy-tailed. Closeness centrality gives a narrower band 2 in time-slice $t_1$ when compared to the original network, which is due to the rewiring process. PageRank, which attributed higher joint community-time scores to nodes in C1 in the original network, now attributes higher scores to nodes in C2 (Fig.~\ref{Fig15}~(k)), and the methods Katz, eigenvector and closeness, which presented slightly scores to C1 in the original network, now attribute slightly higher scores to C2 (Fig.~\ref{Fig15}~(i), (j) and (l)). All methods now capture only one node in band 1 (Fig.~\ref{Fig15}~(m)-(r)), which are kept the same over time. This node is the same for T-ICM, degree, eigenvector, closeness and Katz centralities, and is one of the first ranked nodes in the original network. PageRank, however, attributes the highest score to a different node, which is the same node it attributed the highest score in the original network.

MCC scores are now slightly higher and close to $0.5$ for C2 according to every method, as opposed to slightly higher for C1 as before (except degree centrality, which was higher for C2 in the original network). Eigenvector centrality now also attributes similar MCC scores to both communities, which may be due to the randomisation process decreasing the modularity of the network (modularity is now $4 \times 10^{-4}$ against $0.22$ in the original network). 

\begin{figure}[!ht]
\includegraphics[width = 0.8\textwidth]{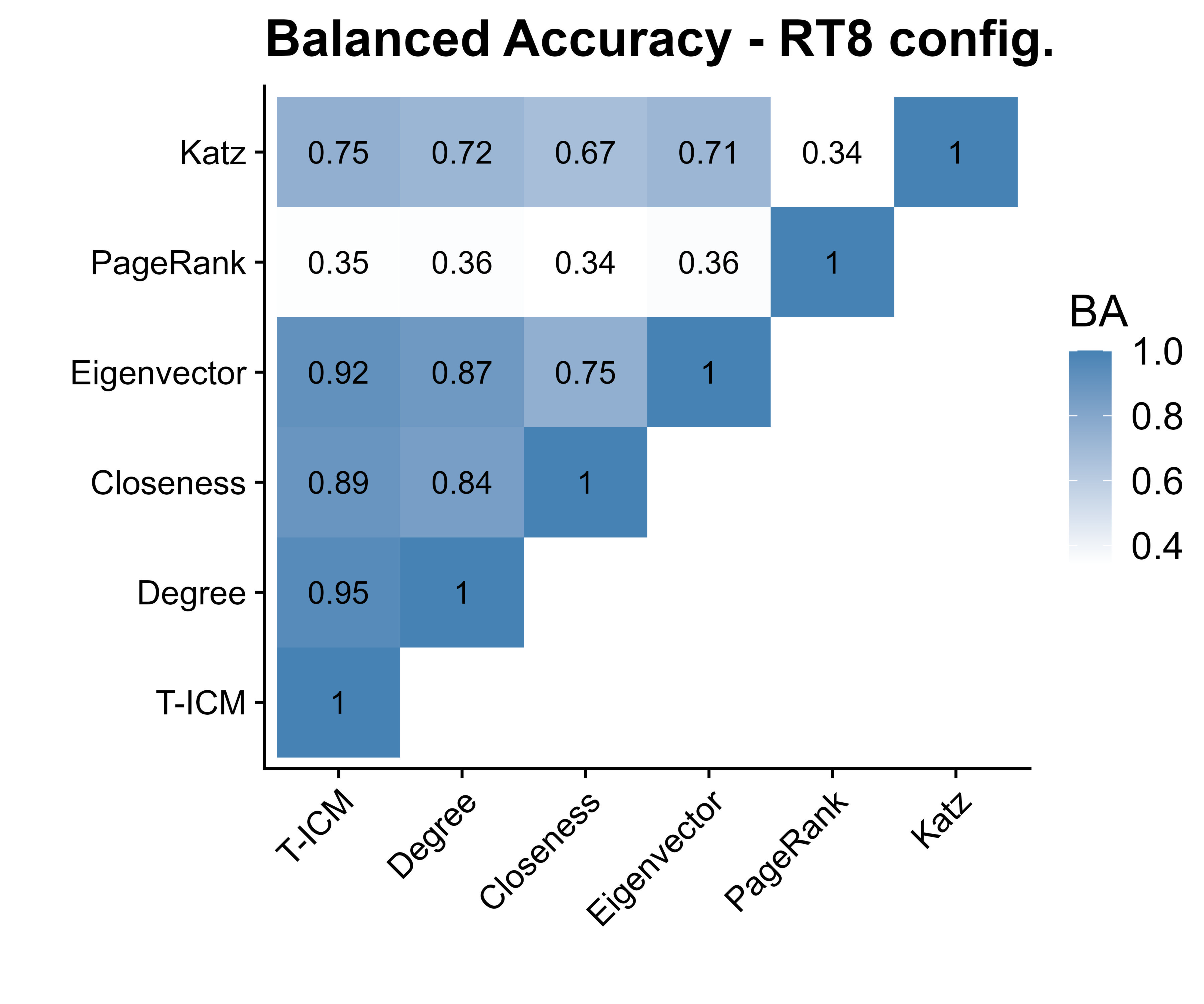}
\caption{{\bf Balanced accuracy between pairs of influence methods in the original RT8 network.} Darker colours represent higher balanced accuracy.}
\label{Fig16}
\end{figure}

For the same reason, the balanced accuracy (Fig.~\ref{Fig16}) of eigenvector centrality increased compared to the values for the original network. The balanced accuracy for Katz, closeness and degree centralities against T-ICM decreased slightly when compared to the values for the original network, however. 

With this analysis, we conclude that by randomising the network structure using the configuration model, the methods are still able to capture the same highest ranked nodes as for the original network, however their overall behaviour and performance are compromised.

\section{Conclusions}

In this paper, we discuss in detail centrality measures that can be used for networks with communities that evolve over time. We started by building synthetic networks with community and bands that allowed us to evaluate the performance of centrality methods in a controlled environment where the influence band of each node in each time-slice was known. We showed that we can successfully aggregate nodes into influence bands (a low-score, a mid-score and a high-score bands), and how to aggregate centrality scores to analyse the influence of communities over time. Additionally, we derived matrices of temporal spread of information that are potentially useful in more theoretical frameworks to compute influence spread in complex networks.

We then studied the influence of communities over time in polarised temporal networks, according to different methods of centrality and influence diffusion. We showed that our modified version of the T-ICM is a good benchmark for centrality methods in this type of network. Using our version of T-ICM we assessed the performance of the centrality methods in a real-world polarised network.

From our analysis, T-ICM and degree centrality perform the best in this setting, as they are able to reliably isolate nodes into their bands. Eigenvector and closeness centrality, however, do not generate the expected results and do not perform well in polarised networks. The temporal eigenvector centrality presents good performance in randomised networks, where modularity is decreased. However, the rank of nodes computed for the randomised network, where we have isolated certain network properties, cannot be deemed as being the same rank for the original network. PageRank performs well in the controlled synthetic networks. However, it does not match the behaviour of our T-ICM benchmark in the more complex setting of our real network. Katz centrality seems to perform better in networks with a more complex degree distribution and communities of different sizes (i.e., BandNet3 and RT8 original) than in simpler networks.

Furthermore, in our studied networks, we observe that the size of the community does not necessarily dictate how influential this community is in the whole network. This requires further investigation. In addition, other clustering methods may be tested, and if the clustering technique chosen affects the results may be investigated. Finally, the optimisation of parameters $\alpha$ and $\varepsilon$ remains an open research opportunity. This would be easy to explore as we already have created the simulations scheme required to do so. 

\section*{Supporting information}

\paragraph*{S1. Randomisation of networks with communities}
\label{S1_Appendix}

Newman et al.~\cite{newman2001random} investigated network properties, such as clustering coefficient, average degree and shortest paths of random networks with arbitrary degree distributions. They showed that the random graph models constructed from real networks perform well in estimating quantities investigated, and in some cases give results of high accuracy. Creating a random network that maintains the degree distribution of the original network is a common practice among network researchers~\cite{kivela2012multiscale, karsai2011small} as it simplifies the network structure while still giving good estimates on the network properties.

The configuration model~\cite{bollobas1980probabilistic} is a flexible and powerful type of random network that may take any degree sequence as we please~\cite{newman2018networks}, where the exact degree of each node is specified through in-stubs --- the number of edges ending on the node --- and out-stubs --- the number of edges starting on the node.  

When it comes to networks with communities, such as the Twitter Irish Abortion Referendum network here studied, various types of stubs must be considered. We need to account for not only in-stubs and out-stubs coming from (going to) the same community --- the in-community stubs --- but also for in-stubs and out-stubs between communities --- the inter-communities stubs. Our Twitter network, which we will be referring to as RT8 network~\cite{pena2025finding}, has two communities, therefore four types of stubs must be considered for each node: the in-community in-stubs and out-stubs, and the inter-communities in-stubs and out-stubs. To create the random network, we first create each community separately, each one containing nodes connected through in-community stubs. We then connect the communities by using the inter-community stubs, where a inter-community out-stub of a node in community 1 is connected to the inter-community in-stub of a node in community 2, and vice-versa. In the following section we will discuss  centrality measures results for the synthetic and real-world networks studied in this paper. 

\section*{Acknowledgements}
This publication has emanated from research jointly funded by Taighde Éireann – Research Ireland under Grant number 18/CRT/6049. For the purpose of Open Access, the authors have applied a CC BY public copyright licence to any Author Accepted Manuscript version arising from this submission.

\bibliography{refs}

\end{document}